# Array Layouts for Comparison-Based Searching[*]

*Paul-Virak Khuong*[†] *and Pat Morin*[‡]

March 14, 2017

Abstract. We attempt to determine the best order and search algorithm to store $n$ comparable data items in an array, $A$, of length $n$ so that we can, for any query value, $x$, quickly find the smallest value in $A$ that is greater than or equal to $x$. In particular, we consider the important case where there are many such queries to the same array, $A$, which resides entirely in RAM. In addition to the obvious sorted order/binary search combination we consider the Eytzinger (BFS) layout normally used for heaps, an implicit B-tree layout that generalizes the Eytzinger layout, and the van Emde Boas layout commonly used in the cache-oblivious algorithms literature.

After extensive testing and tuning on a wide variety of modern hardware, we arrive at the conclusion that, for small values of $n$, sorted order, combined with a good implementation of binary search is best. For larger values of $n$, we arrive at the surprising conclusion that the Eytzinger layout is usually the fastest. The latter conclusion is unexpected and goes counter to earlier experimental work by Brodal, Fagerberg, and Jacob (SODA 2003), who concluded that both the B-tree and van Emde Boas layouts were faster than the Eytzinger layout for large values of $n$. Our fastest C++ implementations, when compiled, use conditional moves to avoid branch mispredictions and prefetching to reduce cache latency.

[*]This research is partially funded by NSERC.
[†]AppNexus, pvk@pvk.ca
[‡]Carleton University, morin@scs.carleton.ca

## Contents

# 1 Introduction

A sorted array combined with binary search represents *the* classic data structure/query algorithm pair: theoretically optimal, fast in practice, and discoverable by school children playing guessing games. Although sorted arrays are *static*—they don't support efficient insertion or deletion—they are still the method of choice for bulk or batch processing of data. Even naïve implementations of binary search execute searches several times faster than the search algorithms of most popular dynamic data structures such as red-black trees.[1]

It would be difficult to overstate the importance of algorithms for searching in a static sorted set. Every major programming language and environment provides a sorting routine, so a sorted array is usually just a function call away. Many language also provide a matching binary search implementation. For example, in C++, sorting is done with std::sort() and the binary search algorithm is implemented in std::lower_bound() and std::upper_bound(). Examples of binary search in action abound; here are two:

1. The Chromium browser code-base calls std::lower_bound() and std::upper_bound() from 135 different locations in a wide variety of contexts, including cookie handling, GUI layout, graphics and text rendering, video handling, and certificate management.[2] This code is built and packaged to make Google Chrome, a web browser that has more than a billion users [21].

2. Repeated binary searches in a few sorted arrays represent approximately 10% of the computation time for the AppNexus real-time ad bidding engine. This engine runs continuously on 1500 machines and handles 4 million requests per second at peak.

However, sorted order is just one possible layout that can be used to store data in an array. Other layouts are also possible and—combined with the right query algorithm— may allow for faster queries. Other array layouts may be able to take advantage of (or be hurt less by) modern processor features such as caches, instruction pipelining, conditional moves, speculative execution, and prefetching.

In this experimental paper we consider four different memory layouts and accompanying search algorithms. The following points describe the scope of our study:

- We only consider array layouts that store $n$ data items in a single array of length $n$.

- The search algorithms used for each layout can find (say) the index of the largest value that is greater than or equal to $x$ for any value $x$. In case $x$ is greater than any value in the array, the search algorithm returns the index $n$.

- We study real (wall-clock) execution time, not instruction counts, branch mispredictions, cache misses, other types of efficiency, or other proxies for real time.

---

[1] For example, Barba and Morin [2] found that a naïve implementation of binary search in a sorted array was approximately three times faster than searching using C++'s stl::set class (implemented as a red-black tree).

[2] https://goo.gl/zpSdXo

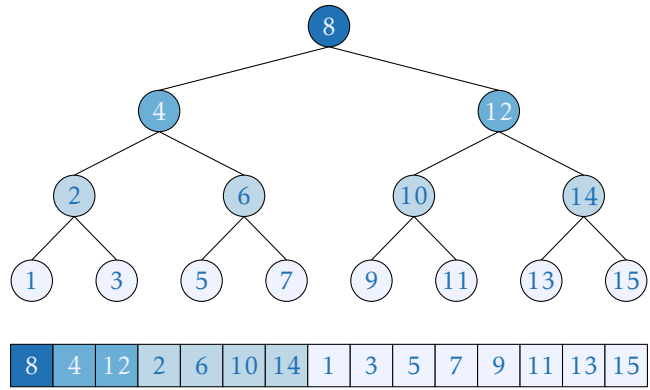

Figure 1: The Eytzinger layout

- We ignore the time required to construct the array layout.[3]

We consider the following four layouts (here and throughout, $\log n = \log_2 n$ denotes the binary logarithm of $n$):

1. Sorted: This is the usual sorted layout, in which the data is stored in sorted order and searching is done using binary search.

2. Eytzinger (see Figure 1): In this layout, the data is viewed as being stored in a complete binary search tree and the values of the nodes in this virtual tree are placed in the array in the order they would be encountered in a left-to-right breadth-first traversal. The search algorithm simulates a search in this implicit binary search tree.

3. Btree (see Figure 2): In this layout, the data is viewed as being stored in a complete $(B+1)$-ary search tree, so that each node—except possibly one leaf—stores $B$ values. The parameter $B$ is chosen so that $B$ data items fit neatly into a single cache line and the nodes of this tree are mapped to array locations in the order they are encountered in a breadth-first search.

4. vEB (see Figure 3): In this, the *van Emde Boas* layout, the data is viewed as being stored in a complete binary tree whose height is $h = \lceil \log(n+1) \rceil - 1$. This tree is laid out recursively: If $h = 0$, then the single node of the tree is stored in $A[0]$. Otherwise, the tree is split into a top-part of height $\lfloor h/2 \rfloor$, which is recursively laid out in $A[0, \ldots, 2^{1+\lfloor h/2 \rfloor} - 2]$. Attached to the leaves of this top tree are up to $2^{1+\lfloor h/2 \rfloor}$ subtrees, which are each recursively laid out, in left-to-right order, starting at array location $A[2^{1+\lfloor h/2 \rfloor} - 1]$.

---

[3] All four layouts can be constructed in $O(n)$ time given a sorted array. Though we don't report construction times in the current paper, they are included in the accompanying data sets. For all layouts it is faster to construct a layout for an array of size $10^8$ than it is to perform $2 \times 10^6$ searches on the resulting layout.

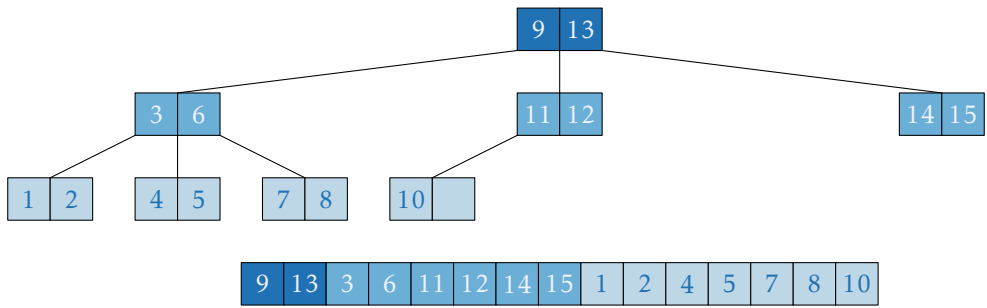

Figure 2: The Btree layout with $B = 2$

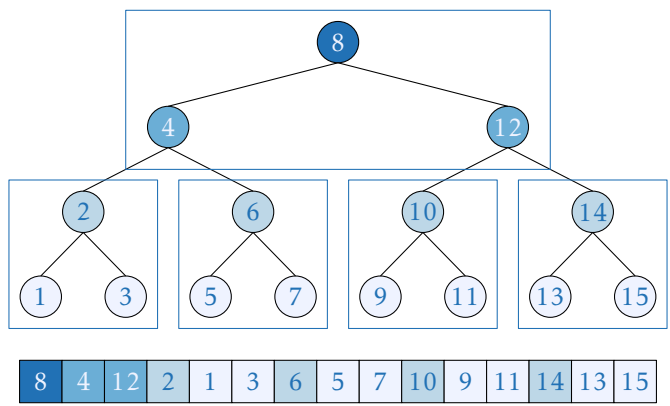

Figure 3: The vEB layout

## 1.1 Related Work

Of the four memory layout/search algorithm pairs we study, the sorted array combined with binary search is, of course, the most well-known and variants of binary search have been studied for more than half a century. Knuth [15, Section 6.2.1] is a good reference.

The idea of implicitly representing a complete binary tree by listing its nodes in breadth-first order goes back half a millenium to Eytzinger [6]. In more modern times, this method was proposed by Williams for an implementation of binary heaps [26].

The implicit Btree layouts we study are a marriage of the Eytzinger layout mentioned above and complete $(B+1)$-ary trees. This layout was studied by Jones [12] and LaMarca and Ladner [16] in the context of implementing heaps. Niewiadomski and Amaral [18] consider an implicit $k$-ary heap implementation in which each node is itself represented using the Eytzinger layout.

The vEB layout was proposed by Prokop [20, Section 10.2] because it has the advantage over the Btree layout of being *cache-oblivious*; the number of cache lines used during a search is within a factor of four of what can be obtained with a B-tree using an optimal choice of $B$.

Storing a static set in an array in order to search it is mainly considered a "solved problem" and, as such, there is little recent work that considers exactly this problem. The lone exception is that of Brodal, Fagerberg, and Jacob [4] who revisited this problem in the context of cache-oblivious algorithms. As part of their work on cache-oblivious search trees, they present experimental results for all four of the layouts we study here. Based on their experiments, they reach the following conclusions:

1. For array lengths smaller than the cache size, the layouts with simplest address arithmetic—sorted and Eytzinger—were the fastest. The Btree layout was next and the vEB layout, with its complicated address arithmetic, was by far the slowest.

2. For array lengths much larger than the cache size, Btree layouts were the fastest, followed by vEB, then Eytzinger, and finally sorted was the slowest.

Between these two extremes, there is an interval where their vEB implementation spends its time catching up to the sorted and Eytzinger implementations.

In a few other cases, searching in a static set comes up as a subroutine within other algorithms. Sanders and Winkel [22] describe a sorting algorithm *Super Scalar Sample Sort (SSSS)* that requires a subroutine for locating values within a sorted sample of the input. They do this efficiently by storing the sample using the Eytzinger layout and show that locating values within this sample can be done using branch-free code (one of the tools we make use of in our implementations).

Sanders and Winkel's implementation of search in the Eytzinger layout is similar to some of ours, in that it uses branch free instructions, but it does not use explicity prefetching (which our code does). In their application, the sample array is small ($k = 256$) so that it fits into even the smallest (fastest) caches, so it does not need to use prefetching.

Their application also performs batched searches in which a large number of elements are being located in this relatively small sample array. They use loop unrolling or software pipelining so that several of these searches can proceed in parallel.

One of the insights of the current paper is that, for smaller array sizes, a branch-free implementation of binary search outperforms Eytzinger search (see Point 2 in the next Section and Figure 12). Thus, is seems that the SSSS implementation could be sped up even further by using branch-free binary search on a sorted array instead of the Eytzinger layout. We have discussed this already with the authors and maintainers of the SSSS code.

Binary search in very small arrays is a subroutine that also comes up in search tree implementations. Kim et al. [14] describe a binary search tree implementation that makes use of (among other things) SIMD instructions to speed up comparison-based searching. In particular, their implementation uses Intel's SSE instructions to perform 3 comparisons in parallel and thus traverse two levels of a binary search tree. It may be possible to use this idea to speed up our implementations of B-trees,[4] though we haven't attempted this because it places more restrictions on the compiler, the hardware, and the supported types of data.

Chen et al. [5] use explicit prefetching to increase the effective cache line width in order to speed up an implementation of Btrees. This is quite far from the current work, but the reason their approach works is that it exploits parallelism in the memory subsystem. It is this effect (combined with speculative execution) that causes some of the surprisingly results in the current paper.

### 1.2 Summary of Results

For readers only interested in an executive summary, our results are the following: For arrays small enough to be kept in L2 cache, the branch-free binary search code listed in Listing 2 is the fastest algorithm (see Figure 11). For larger arrays, the Eytzinger layout, combined with the branch-free prefetching search algorithm in Listing 6 is the fastest general-purpose layout/search algorithm combination (see Figures 10–22).[5] For full experimental data sets on a wide variety of processors, interested readers are directed to this project's webpage.[6]

In more detail, our findings, with respect to a single process/thread executing repeated random searches on a single array, $A$, are summarized as follows:

1. For small values of $n$ (smaller than the L1 cache), branch-free implementations of search algorithms are considerably faster than their branchy counterparts, sometimes by a factor of two.

2. For small values of $n$ (smaller than the L2 cache), a good branch-free implementation of binary search is unbeaten by any other strategy. A branch-free implementation of Eytzinger search is a close second.

---

[4]With the more modern AVX2 instructions, one can even perform 8 comparisons in parallel.
[5]Warning: for consistent performance, the Eytzinger code should use masking as described in Section 5.3.
[6]http://cglab.ca/~morin/misc/arraylayout-v2/

3. For large values of $n$ (larger than the L3 cache), the branch-free implementation of binary search is among the worst of all algorithms, followed closely by the branch-free implementations of Eytzinger search.

4. For large values of $n$ (larger than the L3 cache), the branchy implementations of search algorithms usually perform better than their branch-free counterparts. However, in most cases the difference in performance can be recovered by using explicit prefetching.

5. For large values of $n$, the fastest method is the Eytzinger layout combined with a branch-free search that uses explicit prefetching. More generally, for large values of $n$, the fastest search algorithm for each layout uses a branch-free search combined with explicit prefetching.

6. The standard I/O model of computation [1] is insufficient to explain our results, though a simple extension of this model does explain them. This model suggests two variants of our experiments. Preliminary tests with these variants show that this model has predictive power.

Our conclusions holds across a wide variety of different processors and also hold in a setting in which multiple threads are performing repeated searches on a single array. Our conclusions mostly agree with those of Brodal, Fagerberg and Jacob for small values of $n$, but are completely different for large values of $n$. The reason for the differences is that there are a number of processor architecture considerations—beyond caching—that affect the relative performance of these layouts.

Our Conclusion 1 above supports the choice of a branch-free implementation of the search algorithm in Sanders and Winkel's Super Scalar Sample Sort, though our Conclusion 2 suggests that a good branch-free implementation of a binary search in a sorted array would do slightly better. See, for example, our Figure 12 at the value $n = 2^8$. However, this change would have little impact on the overall running-time of Super Scalar Sample Sort, since this part of the algorithm accounts for only a minor proportion of total execution time [22, Figure 3].

Note that we have been unable to directly compare our implementations to those of Brodal, Fagerberg and Jacob. Although they have kindly made their implementations available to us, we have been unable to use their code for several reasons: (1) their code assume 32-bit processor architectures and fails on 64-bit machines so would require modifications for most of our test machines; (2) their code is C that is generated by a Perl script. This allows them to do many things with C code (that can now be done with C++ facilities, especially templates) but makes it very difficult for us to make even simple changes to their code while maintaining any confidence that the resulting C code will be correct and not drastically different, performance-wise, from their original implementation.

It was only through careful and controlled experimentation with different implementations of each of the search algorithms that we are able to understand how the interactions between processor features such as pipelining, prefetching, speculative execution, and conditional moves affect the running times of the search algorithms. With this understanding, we are able to choose layouts and design search algorithms that perform searches in 1/2 to 2/3 (depending on the array length) the time of the C++ `std::lower_bound()` implementation of binary search (which itself performs searches in 1/3 the time of searching in the `std::set` implemementation of red-black trees).

### 1.3 Outline

The remainder of this paper is organized as follows: Section 2 provides a brief review of modern processor architecture paying particular attention to aspects that affect the results in the current paper. Section 3 describes our implementations of the four memory layouts and experimental results for these implementations. Section 4 proposes a formal model of computation that provides an explanation for our results. Section 5 discusses further experimental results for our implementations. Finally Section 6 summarizes and concludes with directions for further research.

## 2 Processor Architecture Considerations

In this section, we briefly review, at a very high level, the elements of modern processor architecture that affect our findings. For concreteness, we will use numbers from a recent high-end desktop processor, the Intel 4790K [10] with 4 8GB DDR3-1866 memory modules. This processor/RAM combination is also the test machine used for generating the experimental data in Section 3. For a more detailed presentation of this material (though without the running 4790K example), we recommend Patterson's text [19].

### 2.1 CPU

At the highest level, a computer system consists of a processor (CPU) connected to a random access memory (RAM). On the Intel 4790K, the CPU runs at frequency of 4GHz, or $4\times 10^9$ cycles per second. This CPU can execute roughly $4\times 10^9$ instructions per second.[7]

### 2.2 RAM, Latency, and Transfer Rate

The RAM on this system runs at 1866MHz, or roughly $1.866\times 10^9$ cycles per second. This RAM module can transfer 8 bytes per cycle from RAM to the CPU, for a (theoretical) peak transfer rate of $8\times 1.866\times 10^9 \approx 15$GB/s.

Individual memory transfers, however, incur *latency*. The *(first-word) latency* of this RAM module is approximately 10ns: From the time a processor issues a request to read a word of memory from an open row until that word is available is approximately $10^{-8}$ seconds. If the memory is not in an open row (as occurs when this access is far from the previous memory access), latency roughly doubles to $2\times 10^{-8}$ seconds.

Observe that, if the CPU repeatedly reads 4 byte quantities from random locations in RAM, then it receives $1/(2\times 10^8)$ of these per second, for a transfer rate of $4\times(1/2)\times 10^8 = 0.2$GB/s. Note how far this is below this peak transfer rate of 15GB/s.

---

[7]This is only a very rough approximation of the truth; different instructions have different latencies and throughput [8]. Ideal code such as dense linear algebra can sustain 3–4 instructions per cycle, but 0.7 instructions per cycle is normal for typical business code such as online transaction processing [24].

This discrepancy is important: If the CPU is executing instructions that require the contents of memory locations in RAM, and a subsequent instruction cannot begin before the previous instruction completes, then the CPU will not execute more than $(1/2) \times 10^8$ instructions per second; it will waste approximately 79/80 cycles waiting on data from RAM.

When the CPU reads a RAM address, the RAM moves a 64 byte cache line into the CPU. If the processor repeatedly reads cache lines from RAM, this results in a transfer rate of $64/(2 \times 10^{-8}) \approx 3.2\text{GB/s}$. Observe that this is still less than a quarter of the RAM's peak transfer rate.

The key point to take from the preceding discussion is the following: In order to actually achieve a transfer rate close to the theoretical peak transfer rate, the CPU must issue memory read requests before previous requests have finished. This will be important in understanding our experimental results.

## 2.3 Caches

Since reading from RAM is a relatively slow operation, processors use several levels of caches between the processor and the RAM. When the CPU reads a memory location, the entire cache line containing that memory location is loaded into all levels of the CPU cache. Subsequent accesses to memory locations in that cache line then occur with less latency since they can use the data in the CPU caches.

Further complicating matters is the fact that not all caches are fully associative. In a $c$-way associative cache, any memory address, $a$, can only live in one of $c$ cache lines, and in almost all architectures, these cache lines are completely determined by $\lfloor a/B \rfloor \bmod c$. Since $B$ and $c$ are typically powers of 2, this corresponds to $\log c$ bits of $a$ beginning at the $\log B$th least significant bit of $a$.

The Intel 4790K has a 32KB 8-way associative L1 data cache (per core), a 256KB 8-way associative L2 cache (per core), and an 8MB 16-way associative L3 cache (shared among 4 cores). Each of these cache levels is successively slower, in terms of both latency and bandwidth, than the previous level, with L1 being the fastest and L3 being the slowest; but still much faster than RAM.

## 2.4 The Prefetcher

To help achieve peak memory throughput and avoid having the processor stall while waiting on RAM, the CPU includes a prefetcher that analyzes memory access patterns in an attempt to predict future memory accesses. For instance, in simple code such as the following,

```
long sum = 0;
for (int i = 0; i < n; i++)
    sum += a[i];
```

the prefetcher is likely to detect that memory allocated to array a is being accessed sequentially. The prefetcher will then load blocks of a into the cache hierarchy even before they are accessed. By the time the code actually needs the value of a[i] it will already be available in L1/L2/L3 cache.

Prefetchers on current hardware can detect simple access patterns like the sequential pattern above. More generally, they can often detect arithmetic progressions of the form $a, a+k, a+2k, a+3k, \ldots$ and even interleaved arithmetic progressions such as $a, b, a+k, b+r, a+2k, b+2r, a+3k, b+3r, \ldots$. However, current technology does not go much beyond this.

## 2.5 Translation Lookaside Buffer

As part of modern virtual memory systems, the processor has a *translation lookaside buffer (TLB)* that maps virtual memory addresses (visible to processes) to physical memory addresses (addresses of physical RAM). Since a TLB is used for every memory access, it is very fast, and not very large. The TLB organizes memory into fixed-size pages. A process that uses multiple pages of memory will sometimes access a page that is not in the TLB. This is costly, and triggers the processor to walk the *page table* until it finds the appropriate page and then it loads the entry for this page into the TLB.

The Intel 4790K has three data TLBs: The first contains 4 entries for 1GB pages, the second contains 32 entries for 2MB pages, and the third contains 64 entries for 4KB pages. In our experiments—which were done on a dedicated system running few other processes—TLB misses were not a significant factor until the array size exceeded 4GB.

## 2.6 Pipelining, Branch-Prediction, and Predicated Instructions

Executing an instruction on a processor takes several clock cycles, during which the instruction is (1) fetched, (2) decoded, (3) an effective address is read (if necessary), and finally the instruction is (4) executed. Since the entire process takes several cycles, this is arranged in a pipeline so that, for example, one instruction is being executed while the next instruction is reading a memory address, while the next instruction is being decoded, while the next instruction is being fetched.

The Nehalem processor architecture, on which the Intel 4790K is based, has a 20–24 stage processor pipeline [3]. If an instruction does not stall for any other reason, there is still at least a 20–24 cycle latency between the time an instruction is fetched and until the time it completes execution.

Processor pipelining works well provided that the CPU knows which instruction to fetch. Where this breaks down is in code that contains *conditional jump* instructions. These instructions will possibly change the flow of execution based on the result of some previous comparison. In such cases, the CPU does not know in advance whether the next instruction will be the one immediately following the conditional jump or will be the target of the conditional jump. The CPU has two options:

1. Wait until the condition that determines the target of the jump has been tested. In this case, the instruction pipeline is not being filled from the time the conditional jump instruction enters the pipeline until the time the jump condition is tested.

2. Predict whether the jump will occur or not and begin loading the instructions from

the jump target or immediately after the jump, respectively. If the prediction is correct, then no time is wasted. If the prediction is incorrect, then once the jump condition is finally verified, all instructions placed in the pipeline after the conditional jump instruction have to be flushed.

Many processors, including the Intel 4790K, implement the second option and implement some form of *branch predictor* to perform accurate predictions. Branch predictors work well when the condition is highly predictable so that, e.g., the conditional jump condition is almost always taken or almost always not taken.

Most modern processors use some form of two-level adaptive branch-predictor [27] that can even handle second-order statistics, such as conditional jumps that implement loops with a fixed number of iterations. In this case, they can detect conditions such as "this conditional jump is taken $k$ times consecutively and then not taken once." In standard benchmarks, representative of typical work-loads, branch-predictor success rates above 90% and even above 95% are not uncommon [28].

Another useful tool used to avoid branch misprediction (and branches altogether) is the *conditional move* (cmov) family of instructions. Introduced into Intel architectures in 1995 with the Pentium Pro line, these are instructions that move the contents of one register to another (or to memory), but only if some condition is satisfied. They do not change the flow of execution and therefore do not interfere with the processor pipeline.

Conditional moves are a special case of *predicated instructions*—instructions that are only executed if some predicate is true. The Intel IA-64 and ARM CPU architectures include extensive predicated instruction sets.

## 3  The Layouts

In this section, we provide an in-depth discussion of the implementations and performance of the four array layouts we tested.

Throughout this section, we present experimental results. Except where noted otherwise, these results were obtained on the Intel 4790K described in the previous section. In all these experiments, the data consists of 4-byte (32-bit) unsigned integers. In each case, the data stored in the array is the integer set $\{2i+1 : i \in \{0,\ldots,n-1\}\}$ and searches are for uniformly randomly chosen integers in the set $\{0,\ldots,2n\}$. Therefore roughly half the searches were for values in the set and half were for values not in the set. Although the tests reported in this section are for 4-byte unsigned integers, the C++ implementations of the layouts and search algorithms are generic and can be used for any type of data. All of the code and scripts used for our experiments are freely available through github.[8]

Our tests were run under Ubuntu Linux 14.04 with a 3.13 Linux kernel. The compiler used was gcc version 4.8.4 with the command line arguments
```
               -std=c++11 -Wall -O4 -march=native .
```
In all our plots, the array lengths, $n$, are truncated integer powers of $b = 10^{1/10}$. The base $b$ was chosen to provide an even distribution of points when plotted on a logarithmic scale and to avoid array lengths that are powers of 2, which can lead to cache line aliasing caused by limited associativity. The tests shown here were run with 4K TLB pages. Running the same tests with 2M and 1G pages produces the same conclusions.

---

[8]http://dx.doi.org/10.5281/zenodo.31047

## 3.1 Sorted

In this subsection we take special care to understand the performance of two implementations of binary search on a sorted array. Even these two simple implementations of binary search already exhibit some unexpected behaviours on modern processors.

### 3.1.1 Cache-Use Analysis

Here we analyze the cache use of binary search. In this, and all other analyses, we use the following variables:

- $n$ is the number of data items (the length of the array).
- $C$ is the cache size, measured in data items.
- $B$ is the cache-line width, the number of data items that fit into a single cache line.

When we repeatedly execute binary search on the same array, there are two ways the cache helps:

1. (Frequently accessed values) After a large number of searches, we expect to find the most frequently accessed values in the cache. These are the values at the top of the (implicit) binary search tree implemented by binary search. If $n > C/B$, each of these values will occupy their own cache line, so the cache only has room for $C/B$ of these frequently accessed values.

2. (Spatial locality) Once an individual binary search has reduced the search range down to a size less than or equal to $B$, the subarray that remains to be searched will occupy one or two cache lines.

Thus, when we run binary search repeatedly on the same array, the first $\log(C/B)$ comparisons are to cached values and the last $\log B$ comparisons are all to values in the same cache line. Thus, on average, we expect binary search to incur roughly $\log n - \log(C/B) - \log B + 1 = \log n - \log C + 1$ cache misses.

Some cache analyses of binary search ignore spatial locality. For instance, Brodal et al. [4] write "The [sorted] layout has bad performance, probably because no nodes in the top part of the tree share cache lines." In theory, however, the spatial locality in the small subtrees should make up for this.

Another often neglected effect is that of cache associativity, which can adversely effect binary search when the the array length is a large power of 2. In a $c$-way associative cache, the top $\log(n/C)$ levels of the implicit search tree must all share the same $c$ cache lines. If $\log(n/C) > c$, this effectively means that the cache effectively has size only $c$.

On the Intel 4790K, whose 8MB L3 cache can store 2048K cached values, we expect to see a sharp increase in search times when $n$ exceeds $2^{21}$, with each additional level of binary search incurring another L3 cache miss and access to RAM. When we plot search time on a vertical axis versus $n$ on a logarithmic horizontal axis, this shows up as an increase in slope at approximately $n = 2^{21}$.

### 3.1.2 Branchy Binary Search

Our first implementation of binary search is shown in Listing 1. This code implements binary search for a value x the way it is typically taught in introductory computer science courses: It maintains a range of indices {lo,...,hi} and at each stage x is compared to the value a[m] at index, m, in the middle of the search range. The search then either finds x at index m (if x = a[m]) or continues searching on one of the ranges {lo,...,m} (if x < a[m]) or {m+1,...,hi} (if x > a[m]). Since the heart of this algorithm is the three-way branch inside the `while` loop, we call this implementation *branchy binary search*.

```cpp
template<typename T, typename I>
I sorted_array<T,I>::branchy_search(T x) const {
    I lo = 0;
    I hi = n;
    while (lo < hi) {
        I m = (lo + hi) / 2;
        if (x < a[m]) {
            hi = m;
        } else if (x > a[m]) {
            lo = m+1;
        } else {
            return m;
        }
    }
    return hi;
}
```

Listing 1: Source code for branchy binary search.

Figure 4 shows the running time of $2 \times 10^6$ searches for values of $n$ ranging from 1 to $2^{30}$. As the preceding cache analysis predicts, there is indeed a sharp increase in slope that occurs at around $n = 2^{21}$. To give our results a grounding in reality, this graph also shows the running-time of the `stl::lower_bound()` implementation—The C++ Standard Template Library implementation of binary search. Our branchy implementation and the `stl::lower_bound()` implementation perform nearly identically.

If we consider only values of $n$ up to $2^{21}$, shown in Figure 5, we see an additional change in slope at $n = 2^{16}$. This is the same effect, but at the L2/L3 cache level; the 4790K has a 256KB L2 cache capable of storing $64K = 2^{16}$ data items. Each additional level of binary search beyond that point incurs an additional L2 cache miss and an access to the

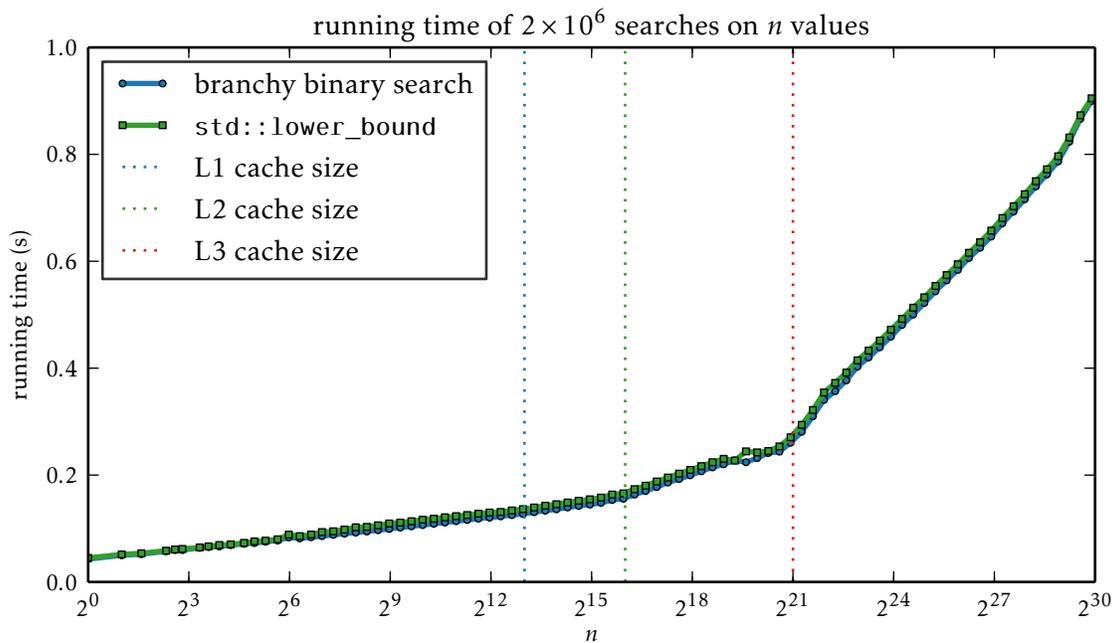

Figure 4: The running time of branchy binary search and stl::lower_bound().

L3 cache.

### 3.1.3 Branch-Free Binary Search

Readers with experience in micro-optimizing code will see that, for modern desktop processors, the code in Listing 1 can be optimized substantially. There are two problems with this code:

1. Inside the code is a three-way if statement whose execution path is highly unpredictable. Each of the first two code blocks has a close to 50% chance of being executed. The branch-predictor of a pipelined processor is forced to guess which of these blocks will occur and load the instructions from this block into the pipeline. When it guesses incorrectly (approximately 50% of the time), the entire pipeline must be flushed and the instructions for the other branch loaded.

2. The number of iterations of the outer loop is hard to predict. The loop may terminate early (because x was found). Even when searching for a value x that is not present, unless $n$ has the form $2^k - 1$, the exact number of iterations is different for different values of x. This implies that the branch predictor will frequently mispredict termination or non-termination of the loop, incurring the cost of another pipeline flush per search.

Listing 2 shows an alternative implementation of binary search that attempts to alleviate both problems described above. (This code implements a variant of Knuth's Algorithm U (Uniform Binary Search) [15, Section 6.2.1].) In this implementation, there is no early termination and, for a given array length n, the number of iterations is fixed (because the value of n always decreases by `half` during each iteration of the loop). Therefore, when this method is called repeatedly on the same array, a good branch-predictor will very quickly learn the number of iterations of the `while` loop, and it will generate no further branch mispredictions.

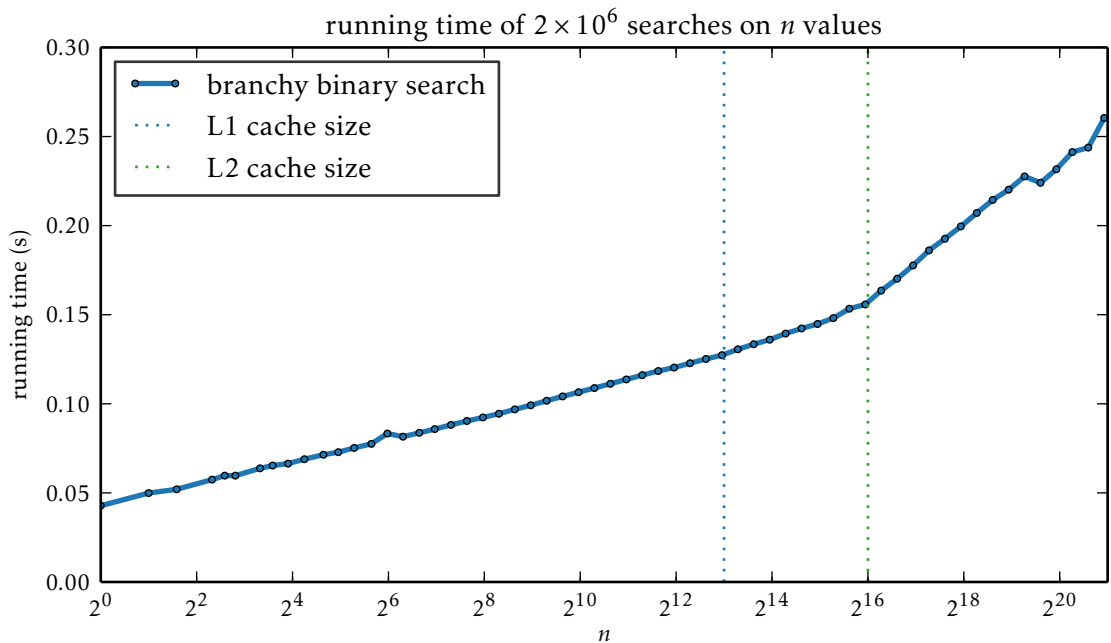

Figure 5: The running time of branchy binary search when all data fits into L3 cache.

```
template<typename T, typename I>
I sorted_array<T,I>::_branchfree_search(T x) const {
    const T *base = a;
    I n = this->n;
    while (n > 1) {
        const I half = n / 2;
        base = (base[half] < x) ? &base[half] : base;
        n -= half;
    }
    return (*base < x) + base - a;
}
```

Listing 2: Source code for branch-free binary search.

In the interior of the `while` loop, there is only one piece of conditional code, which occurs in Line 7. For readers unfamiliar with C's choice operator, this code is equivalent to `if (base[half] < x) base = &base[half]`, so this line either reassigns the value of base (if base[half] < x) or leaves it unchanged. The compiler implements this using a *conditional move* (cmov) instruction so that there is no branching within the while loop. For this reason, we call this *branch-free binary search*.

The use of conditional move instructions to replace branching is a topic of heated debate (see, e.g., [23]). Conditional move instructions tend to use more clock cycles than traditional instructions and, in many cases, branch predictors can achieve prediction accuracies exceeding 95%, which makes it faster to use a conditional jump. In this particular instance, however, the branch predictor will be unable to make predictions with accuracy exceeding 50%, making a conditional move the best choice. The resulting assembly code, shown in Listing 3 is very lean. The body of the `while` loop is implemented by Lines 8–15 with the conditional move at Line 12.

```
    .cfi_startproc
    movq    8(%rdi), %rdx       ; move n into rdx
    movq    (%rdi), %r8         ; move a into r8
    cmpq    $1, %rdx            ; compare n and 1
    movq    %r8, %rax           ; move base into rax
    jbe     .L2                 ; quit if n <= 1
.L3:
    movq    %rdx, %rcx          ; put n into rcx
    shrq    %rcx                ; rcx = half = n/2
    leaq    (%rax,%rcx,4), %rdi ; load &base[half] into rdi
    cmpl    %esi, (%rdi)        ; compare x and base[half]
    cmovb   %rdi, %rax          ; set base = &base[half] if x > base[half]
    subq    %rcx, %rdx          ; n = n - half
    cmpq    $1, %rdx            ; compare n and 1
    ja      .L3                 ; keep going if n > 1
.L2:
    cmpl    %esi, (%rax)        ; compare x to *base
    sbbq    %rdx, %rdx          ; set dx to 00..00 or 11...11
    andl    $4, %edx            ; set dx to 0 or 4
    addq    %rdx, %rax          ; add dx to base
    subq    %r8, %rax           ; compute base - a (* 4)
    sarq    $2, %rax            ; (divide by 4)
    ret
    .cfi_endproc
```

Listing 3: Compiler-generated assembly code for branch-free binary search.

Figure 6 compares the performance of the branchy and branch-free implementations of binary search for array sizes ranging from 1 to $2^{16}$. As expected, the branch-free code is much faster. After accounting for the overhead of the testing harness, the branch-free search is approximately twice as fast for $n = 2^{16}$. Running the two using the perf performance monitoring tool explains why. With $2 \times 10^6$ searches on an array of length $n = 60,000$, the branchy code executes $101 \times 10^3$ branches and mispredicts 17.3% of these. The branch-free code executes $57 \times 10^3$ branches and mispredicts only 0.05% of these.

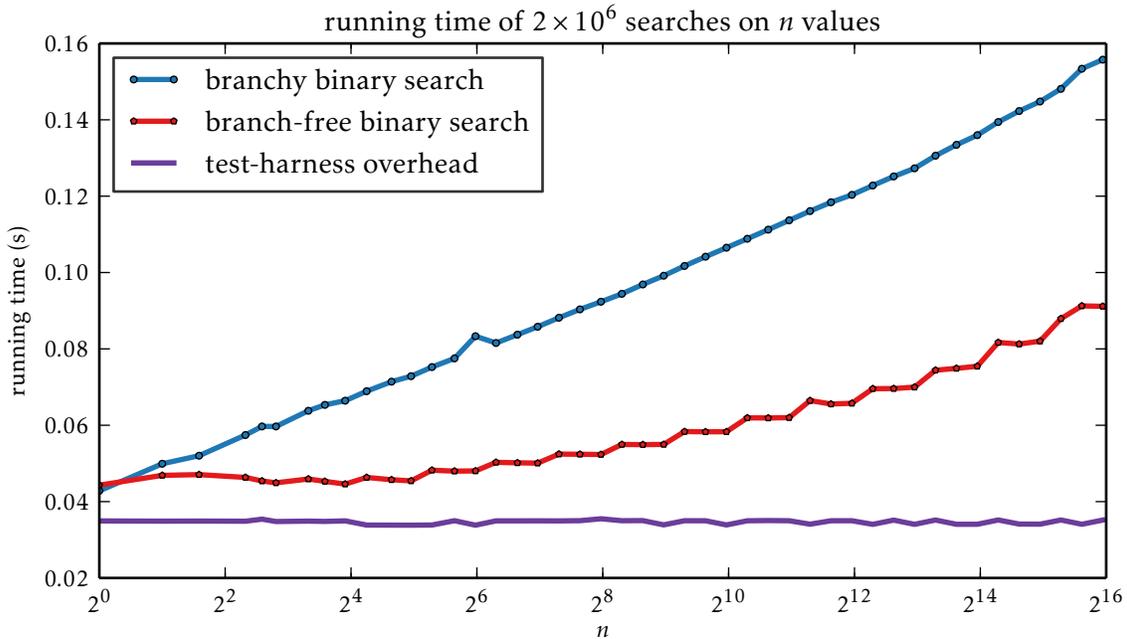

Figure 6: The running times of branchy binary search versus branch-free binary search when all data fits into L2 cache.

However, for larger values of $n$ (shown in Figure 7) the situation changes. For $n > 2^{16}$, the gap begins to narrow slowly until $n > 2^{21}$, at which point it narrows more quickly. By the time time $n$ exceeds $2^{22}$, the branch-free code is slower and the gap between the two continues to widen from this point onward.

### 3.1.4 Branchy Code, Speculative Execution, and Implicit Prefetching

The reason for the change in relative performance between branchy and branch-free binary search was not immediately obvious to us. After some experimentation, we discovered it comes from the interplay between branch prediction and the memory subsystem. In the branchy code, the branch-predictor makes a guess at which branch will be taken and is correct approximately half the time. An incorrrect guess causes a costly pipeline flush. However, a correct guess results in the memory subsystem starting to load the array location needed during the next iteration of the `while` loop.

For $n < 2^{16}$, the entire array fits into L2 cache, and the costs of pipeline flushes exceed any savings obtained by correct guesses. However, for larger $n$, each correct guess by the branch-predictor triggers an L2 (in the range $2^{16} < n < 2^{21}$) or an L3 (for $n > 2^{21}$) cache miss sooner than it would otherwise. The costs of these cache misses are greater than the costs of the pipeline flushes, so eventually, the branch-free code loses out to the branchy code.

Since this was not our initial explanation, and since it was not obvious from the beginning, we gathered several pieces of evidence to support or disprove this hypothesis.

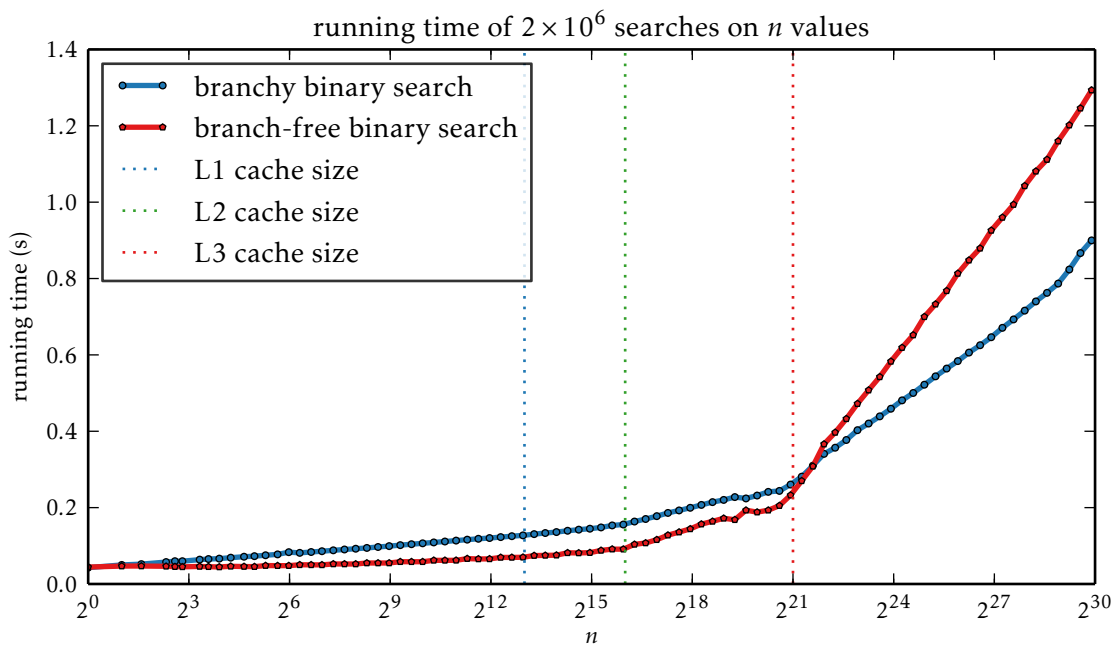

Figure 7: The running times of branchy binary search versus branch-free binary search for large values of $n$.

1. Assembly-level profiling showed that, for large $n$, the branch-free code spends the vast majority of its time loading from memory (Line 11, of Listing 3). This is because the register, `rdi`, containing the memory address to load is the target of a conditional move (Line 12). This conditional move has not completed because it is still waiting on the results of the previous comparison, so execution stalls at this point.

2. We measured L3 cache accesses using `perf`.[9] With $n = 10^8$, the branchy code performs $106 \times 10^6$ L3 cache accesses for $2 \times 10^6$ searches. The branch-free code performs only $33 \times 10^6$ L3 cache accesses for the same test. This difference is caused by speculative execution in the branchy code; each mispredicted branch loads a memory location that is not loaded by the branch-free code.

3. Another possible explanation for our results is that the hardware prefetcher is, for some reason, better able to handle the branchy code than the branch-free code. This seems unlikely, since the memory access patterns for both version are quite similar, and probably too complicated for a hardware prefetcher to predict. Nevertheless, we ruled out this possibility by disabling the hardware prefetcher and running the tests again.[10] The running-times of the code were unchanged by disabling prefetching.

4. We implemented a version of the branch-free code that adds explicit prefetching. At

---

[9]For this, we used the command `perf stat -e r4f2e` to obtain the number of lowest level cache references [11, Table 18-1].

[10]Prefetching was disabled using the `wrmsr` utility with register number 0x1a4 and value 0xf. This disables all prefetching [9].

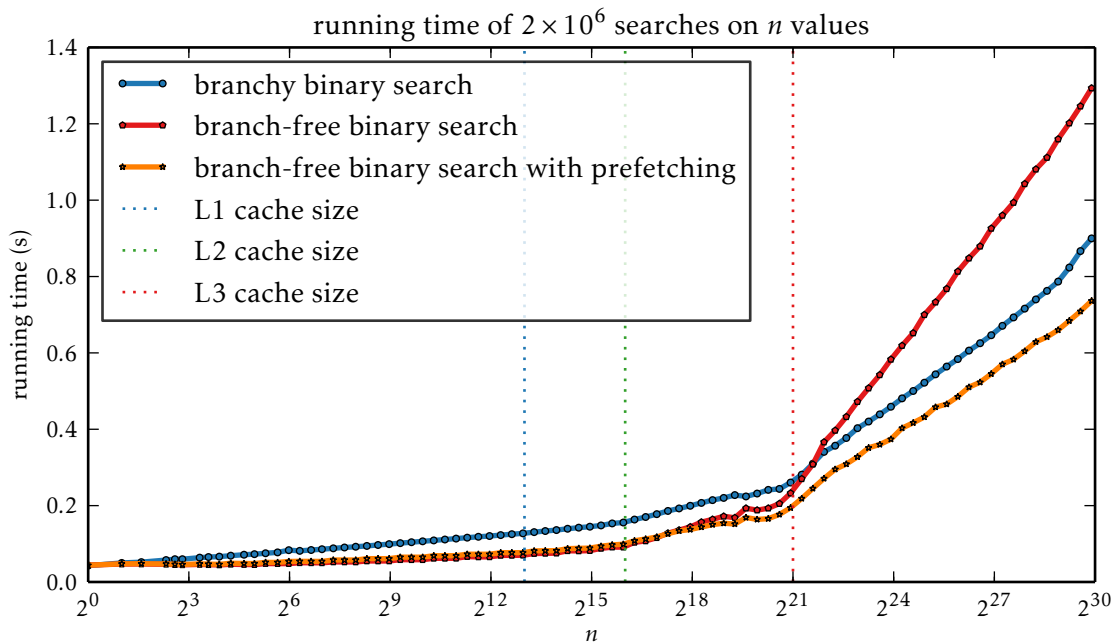

Figure 8: Branch-free binary search with explicit prefetching is competitive for small values of $n$ and a clear winner for large values of $n$.

the top of the while loop, it prefetches array locations a[half/2] and a[half+half/2] using gcc's __builtin_prefetch() builtin, which translates into the x86 prefetch0 instruction. The performance of the resulting code, shown in Figure 8 is consistent with our hypothesis. The code is nearly as fast as the branch-free code for small values of $n$, but tracks (and even improves) the performance of the branchy code for larger values of $n$.

Note that this code actually causes the memory subsystem to do more work, and consumes more memory bandwidth since, in general it loads two cache lines when only one will be used. Nevertheless it is faster because the memory bandwidth is more than twice the cache line width divided by the memory latency.

5. We ran our code on an Atom 330 processor that we had available. This low-power processor was designed to minimize watts-per-instruction so it does not do any form of speculative execution, including branch prediction. The results, which are shown in Figure 9, are consistent with our hypothesis. The branch-free code is faster than the branchy code across the full range of values for $n$. This is because, on the Atom 330, branches in the branchy code result in pipeline stalls that do not help the memory subsystem predict future memory accesses.

From our study of binary search, we conclude the following lesson about the (very important) case where one is searching in a sorted array:

**Lesson 1.** For searching in a sorted array, the fastest method uses branch-free code with explicit prefetching. This is true both on modern pipelined processors that use branch prediction and on traditional sequential processors.

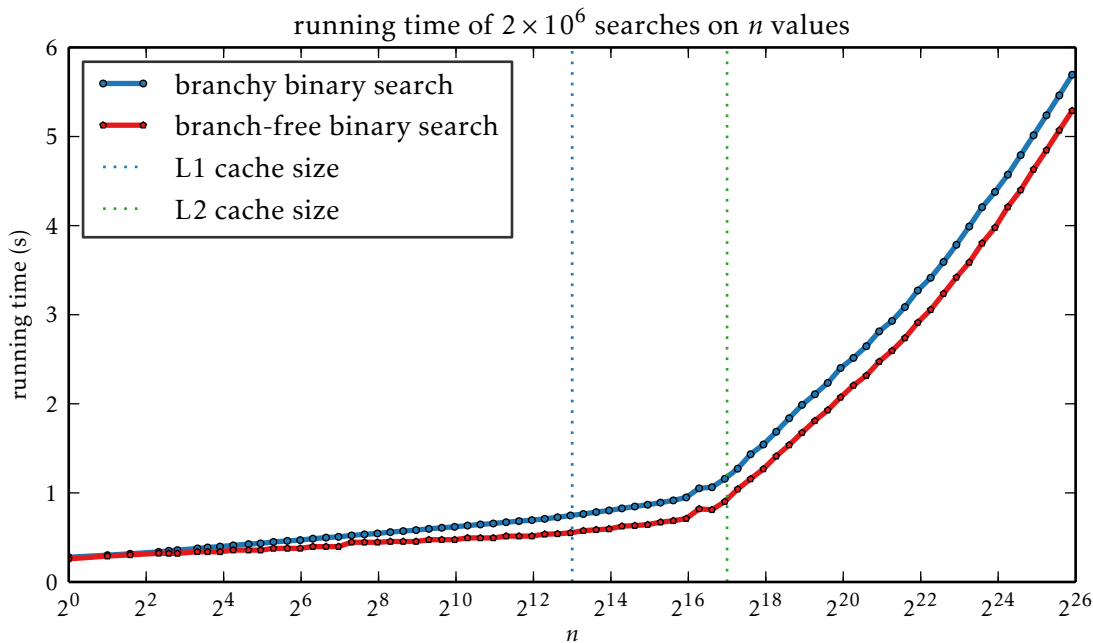

Figure 9: Binary search running-times on the Atom 330. The Atom 330 has a 512KB L2 cache that can store $2^{17}$ 4-byte integers and does not perform speculative execution of any kind.

We finish our discussion of binary search by pointing out one lesser-known caveat: If $n$ is large and very close to a power of 2, then all three variants of binary search discussed here will have poor cache utilization. This is caused by cache-line aliasing between the elements at the top levels of the binary search; in a $c$-way associative cache, these top-level elements are all restricted to use $O(c)$ cache lines. This problem is observed in the experimental results of Brodal et al. [4, Section 4.2] and examined in detail by the first author [13], who also suggests efficient workarounds. Our own experiments mostly sidestep this issue by our choice of array lengths.

### 3.2 Eytzinger

Recall that, in the Eytzinger layout, the data is viewed as being stored in a complete binary search tree and the values of the nodes in this virtual tree are placed in the array in the order they would be encountered in a left-to-right breadth-first traversal. A nice property of this layout is that it is easy to follow a root-to-leaf path in the virtual tree: The value of the root of the virtual tree is stored at index 0, and the values of left and right children of the node stored at index $i$ are stored at indices $2i + 1$ and $2i + 2$, respectively.

### 3.2.1 Cache-Use Analysis

With respect to caching, the performance of searches in an Eytzinger array should be comparable to that of binary searching in a sorted array, but for slightly different reasons.

When performing repeated random searches, the top levels of the (virtual) binary tree are accessed most frequently, with a node at depth $d$ being accessed with probability roughly $1/2^d$. Since nodes of the virtual tree are mapped into the array in breadth-first order, the top levels appear consecutively at the beginning of the array. Thus, we expect the cache to store, roughly, the first $C$ elements of the array, which correspond to the top $\log C$ levels of the virtual tree. As the search proceeds through the top $\log C$ levels, it hits cached values, but after $\log C$ levels, each subsequent comparison causes a cache miss. This results in a total of roughly $\log n - \log C$ cache misses, just as in binary search.

### 3.2.2 Branchy Eytzinger Search

A branchy implementation of search in an Eytzinger array, shown in Listing 4, is nearly as simple as that of binary search, with only one slight complication. When branchy binary search completes, the variable `hi` stores the return value. In Eytzinger search, we must decode the return value from the value of `i` at the termination of the `while` loop. Since we haven't seen this particular bit-manipulation trick before, we explain it here.

```
template<typename T, typename I>
I eytzinger_array<T,I>::branchy_search(T x) const {
    I i = 0;
    while (i < n) {
        if (x < a[i]) {
            i = 2*i + 1;
        } else if (x > a[i]) {
            i = 2*i + 2;
        } else {
            return i;
        }
    }
    I j = (i+1) >> __builtin_ffs(~(i+1));
    return (j == 0) ? n : j-1;
}
```

Listing 4: Branchy implementation of search in an Eytzinger array.

In the Eytzinger layout, the $2^d$ nodes of the virtual tree at depth $d$ appear consecutively, in left-to-right order, starting at index $\sum_{k=0}^{d-1} 2^k = 2^d - 1$. Therefore, an array position $i \in \{2^d - 1, \ldots, 2^{d+1} - 2\}$ corresponds to a node, $u$, of the virtual tree at depth $d$. Furthermore, the binary representation, $b_w, b_{w-1}, \ldots, b_0$, of $i+1$ has a nice interpretation:

- The bits $b_w = b_{w-1} = \cdots = b_{d+1} = 0$, since $i + 1 < 2^{d+1}$.

- The bit $b_d = 1$ since $i + 1 \geq 2^d$.

- The bits $b_{d-1}, \ldots, b_0$ encode the path, $u_1, \ldots, u_d$, from the root of the virtual tree to $u = u_d$: For each $k \in \{1, \ldots, d-1\}$, if $u_{k+1}$ is the left (respectively, right) child of $u_k$, then $b_{d-k} = 0$ (respectively, $b_{d-k} = 1$).

Therefore, the position of the highest order 1-bit of $i+1$ tells us the depth, $d$, of the node $u$ and the bits $b_{d-1}, \ldots, b_0$ tell us the path to $u$. This is true even if $i \geq n$, in which case we can think of $u$ as an external node of the virtual tree (that stores no data). To answer a query, we are interested in the last node, $w$, on the path to $u$ at which the path proceeded to a left child. This corresponds to the position, $r$, of the lowest order 0 bit in the binary representation of $i + 1$. Furthermore, by right-shifting $i + 1$ by $r + 1$ positions we obtain an integer $j$ such that:

1. $j - 1$ is the index of $w$ in the array (if $w$ exists), or

2. $j = 0$ if $w$ does not exist (because the path to $u$ never proceeds from a node to its left child).

The latter case occurs precisely when we search for a value x that is larger than any value in the array. The last two lines of code in Listing 4 extract the value of $j$ from the value of $i$ and convert this back into a correct return value. The only non-standard instruction used here is gcc's `__builtin_ffs` builtin that returns one plus the index of the least significant one bit of its argument. This operation, or equivalent operations, are supported by most hardware and C/C++ compilers [25].

### 3.2.3 Branch-Free Eytzinger Search

A branch-free implementation of search in an Eytzinger array—shown in Listing 5—is also quite simple. Unfortunately, unlike branch-free binary search, the branch-free Eytzinger search is unable to avoid variations in the number of iterations of the while loop, since this depends on the depth (in the virtual tree) of the leaf that is reached when searching for x. When $n = 2^h - 1$ for some integer $k$, then all leaves have the same depth but, in general, there will be leaves of depths $h$ and depth $h - 1$, where $h = \lceil \log(n+1) \rceil - 1$ is the height of the virtual tree.

The performance of the branchy and branch-free versions of search in an Eytzinger array is shown in Figure 10. Branch-free Eytzinger search very closely matches the performance of branch-free binary search. As expected, there is an increase in slope at $n = 2^{16}$ and $n = 2^{21}$ since these are the number of 4-byte integers that can be stored int the L2 and L3 caches, respectively.

However, the performance of the branchy Eytzinger implementation is much better than expected for large values of $n$. Indeed, it is much faster than the branchy implementation of binary search, and even faster than our fastest implementation of binary search.

```
template<typename T, typename I>
I eytzinger_array<T,I>::branchfree_search(T x) const {
    I i = 0;
    while (i < n) {
        i = (x <= a[i]) ? (2*i + 1) : (2*i + 2);
    }
    I j = (i+1) >> __builtin_ffs(~(i+1));
    return (j == 0) ? n : j-1;
}
```

Listing 5: Branch-free implementation of search in an Eytzinger array.

### 3.2.4 Why Eytzinger is so Fast

Like branchy binary search, the branchy implementation of search in Eytzinger array is so much faster than expected because of the interaction between branch prediction and the memory subsystem. However, in the case of the Eytzinger layout, this interaction is much more efficient.

To understand why the branchy Eytzinger code is so fast recall that, in the Eytzinger layout, the nodes are laid out in breadth-first search order, so the two children of a node occupy consecutive array locations $2i$ and $2i+1$. More generally, for a virtual node, $u$, that is assigned index $i$, $u$'s $2^\ell$ descendants at distance $\ell$ from $u$ are consecutive and occupy array indices $2^\ell i + 2^\ell - 1, \ldots, 2^\ell i + 2^{\ell+1} - 1$.

In our working example of 4-byte data with 64-byte cache lines, the sixteen great-great grandchildren of the virtual node stored at position $i$ are stored at array locations $16i + 15, \ldots, 16i + 30$. Assuming that the first element of the array is the first element of a cache line, this means that, of those 16 descendants, 15 are contained in a single cache line (the ones stored at array indices $16i + 16, \ldots, 16i + 30$.)

With the Intel 4790K's 20–24 cycle pipeline, instructions in the pipeline can be several iterations ahead in the execution of the `while` loop, which only executes 5–6 instructions per iteration. The probability that the branch-predictor correctly guesses the execution path of four consecutive iterations is only about 1/16, and a flush of (some of) the pipeline is likely. However, even if the branch predictor does not guess the correct execution path, it most likely (with probability 15/16) guesses an execution path that loads the correct cache line. Even though the branch predictor loads instructions that will likely never be executed, their presence in the pipeline causes the memory subsystem to begin loading the cache line that will eventually be needed.

Knowing this, there are two optimizations we can make:

1. We can add an explicit prefetch instruction to the branch-free code so that it loads the correct cache line. The resulting code is shown in Listing 6.

2. We can align the array so that a[0] is the second element of a cache line. In this way, all 16 great-great grandchildren of a node will always be stored in a single cache line.

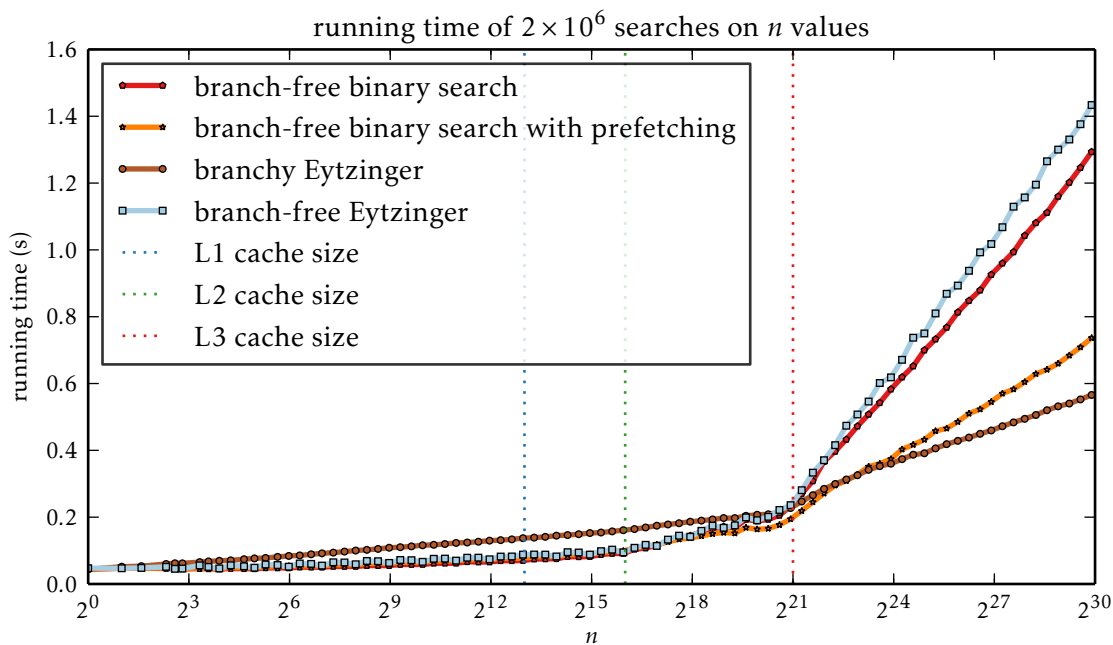

Figure 10: The performance of branchy and branch-free Eytzinger search.

The results of implementing both of these optimizations are shown in Figure 11.[11] With these optimizations, the change in slope at the L2 cache limit disappears; four iterations of the search loop is enough time to prefetch data from the L3 cache. Once the L3 cache limit is reached, the behaviour becomes similar to the branchy implementation, but remains noticeably faster, since the branch-free code does not cause pipeline flushes and always prefetches the correct cache line 4 levels in advance.

Figure 12 compares Eytzinger search with binary search for small values of $n$ ($n \leq 2^{16}$). At this scale the branch-free Eytzinger code is very close, but not quite as fast as the best binary search implementations.

Something else evident in Figure 11 is a "bumpiness" of the Eytzinger runningtimes as a function of $n$. This bumpiness is caused by branch mispredictions during the execution of the while loop. The search times are especially low when $n$ is close to a power of two because in this case nearly all searches perform the same number of iterations of the while loop (since the underlying tree is a complete binary tree of height $h$ with nearly all leaves at same level $h$). The branch predictor quickly learns the value of $h$ then correctly predicts the final iteration. On the other hand, when $n$ is far from a power of 2, the number of iterations is either $h$ or $h-1$, each with non-negligible probability, and the branch predictor is unable to accurately predict the final iteration of the while loop. For example, measuring with perf when $n = 2^{11}$ or $n = 2^{12}$, we find that the branch-free Eytzinger search has a branch-misprediction rate of 0.04% when performing $2 \times 10^7$

---

[11] Although Figure 11 shows the results of implementing both these optimizations, we did test them individually. Realigning the array gives in only a small improvement in performance; the real improvements come from combining branch-free code and explicit prefetching.

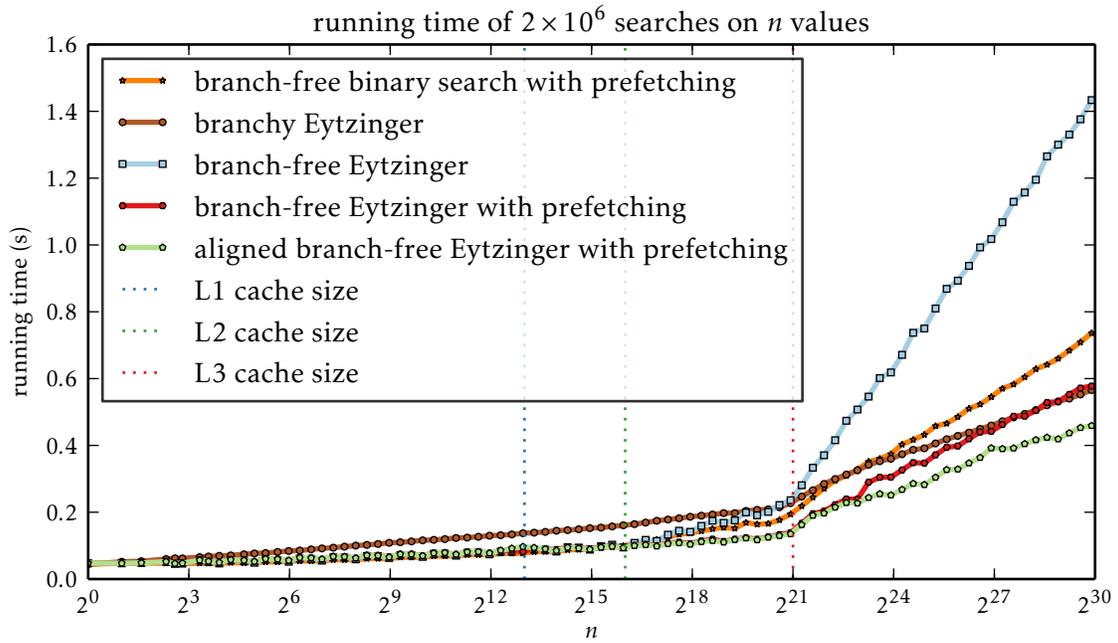

Figure 11: The performance of branch-free Eytzinger with prefetching.

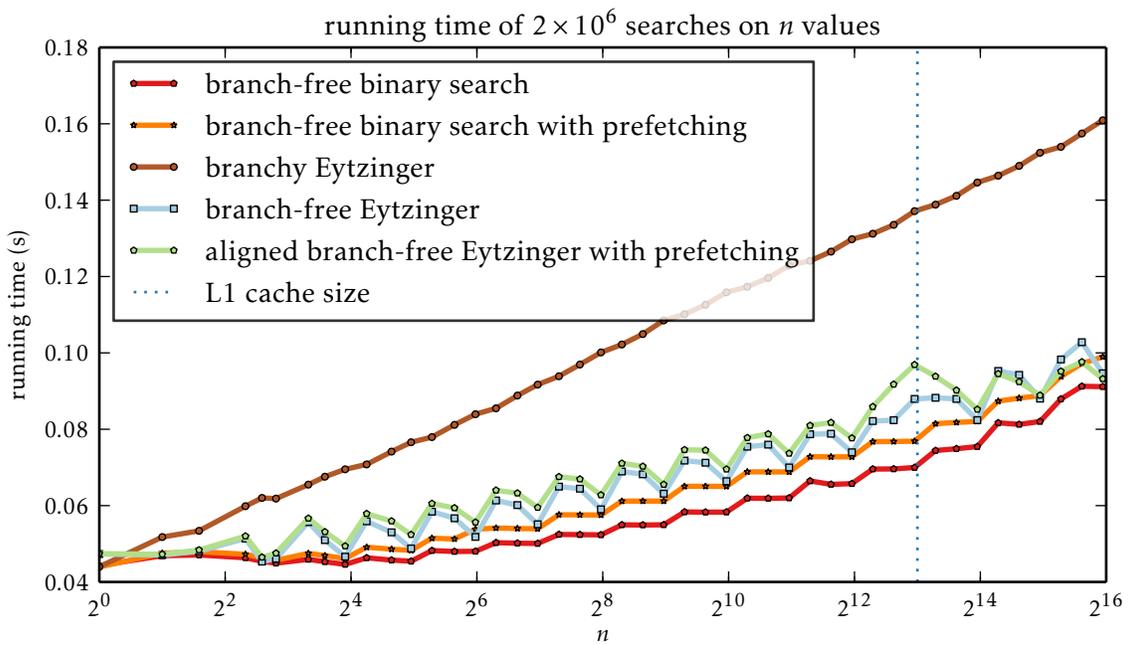

Figure 12: Eytzinger versus binary search for small values of $n$.

```
template<typename T, typename I>
I eytzinger_array<T,I>::branchfree_search(T x) const {
    I i = 0;
    while (i < n) {
        __builtin_prefetch(a+(multiplier*i + offset));
        i = (x <= a[i]) ? (2*i + 1) : (2*i + 2);
    }
    I j = (i+1) >> __builtin_ffs(~(i+1));
    return (j == 0) ? n : j-1;
}
```

Listing 6: Branch-free prefetching implementation of search in an Eytzinger array. (The value of `multiplier` in this code is the cache line width, $B$, and the value of `offset` is $\lfloor 3B/2 \rfloor - 1$.)

searches. When $n = (2^{11} + 2^{12})/2$, the branch misprediction rate is 1.43%.

**Lesson 2.** For the Eytzinger layout, branch-free search with explicit prefetching is the fastest search implementation. For small values of $n$, this implementation is competitive with the fastest binary search implementations. For large values of $n$, this implementation outperforms all versions of binary search by a wide margin.

### 3.2.5 A Mixed Eytzinger/Sorted Layout

We conclude this section with a discussion of a mixed layout suggested by the cache-use analyses of binary and Eytzinger search. Conceptually, this layout implements a binary search tree (laid out as an Eytzinger array) whose leaves each contain blocks of $B$ values (stored in sorted order).

More specifically, the first part of the array contains values $x_0, \ldots x_{2^h-2}$ stored using the Eytzinger layout and the second part of the array contains values $y_0, \ldots, y_{n-2^h}$ in sorted order. Here, the value of $h$ is the minimum integer such that $2^h - 1 + B2^h \geq n$ and the values in the first part of the array are chosen so that

$$y_0 < \cdots < y_{B-1} < x_0 < y_B < \cdots < y_{2B-1} < x_1 < \cdots$$

In this way, a search in the first part of the array usually narrows down the answer to a block of $B$ consecutive values in the second part of the array, and these values are stored in a single cache line. Using this layout, one can store up to $BC$ values and a search will incur only a single cache miss. (The Eytzinger layout lives entirely in the cache of size $C$ and reduces the problem to a binary search in a single cache line of size $B$.)

Both the search in the Eytzinger array and the sorted block can be implemented using fast branch-free code so, in theory, this layout should be faster than either the Eytzinger or the sorted layout. This is true: In Figure 13 we see that this code effectively increases the cache size when compared to the branch-free Eytzinger code. Though there is an increase in slope at $n = 2^{21}$, it does not become as severe as the Eytzinger code until $n = 2^{25}$. (In Figure 13 both implementations are branch-free and neither use prefetching.)

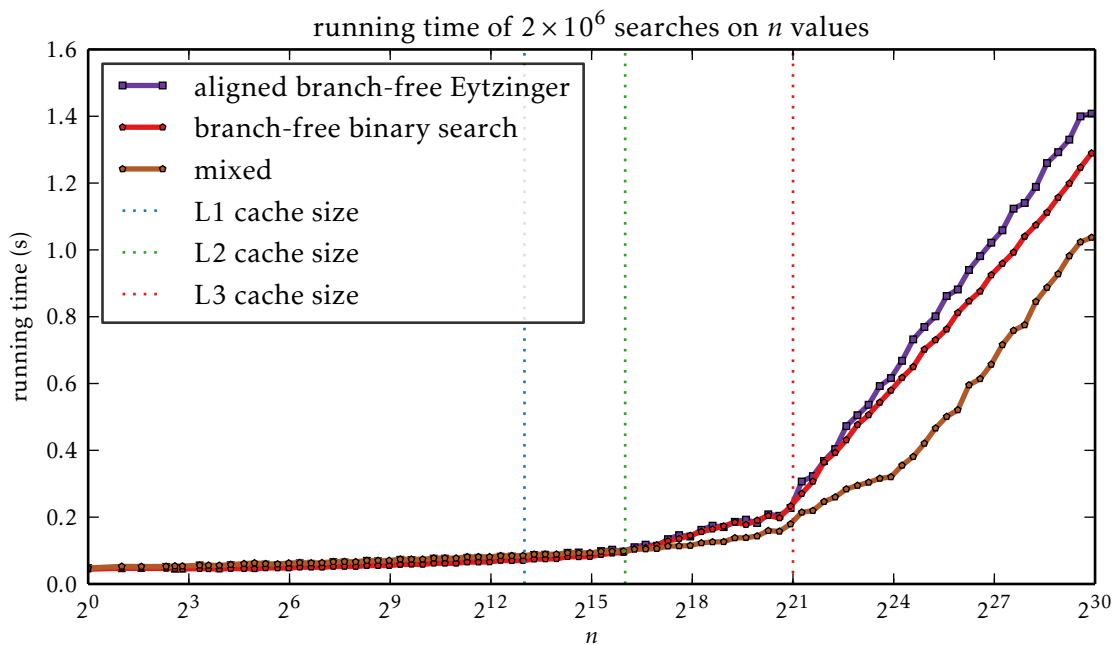

Figure 13: Performance of the branch-free mixed layout (without prefetching).

However, this mixed layout is unable to beat branch-free Eytzinger search with explicit prefetching. This is true even when the mixed layout uses prefetching in the Eytzinger part of its search (which has the side-effect of prefetching the correct block for the second part of the search). Figure 14 compares branch-free Eytzinger with prefetching and a prefetching version of the mixed layout. The two layouts are virtually indistinguishable with respect to performance.

**Lesson 3.** Although the mixed layout looks promising, it is unable to beat the performance of branch-free Eytzinger with prefetching and is considerably more complicated to implement.

### 3.3 Btree

Recall that the $(B+1)$-tree layout simulates search on a $(B+1)$-ary search tree, in which each node stores $B$ values. The nodes of this $(B+1)$-ary tree are laid out in the order they are encountered in a breadth-first search. The Eytzinger layout is a special case of the Btree layout in which $B = 1$. As with the Eytzinger layout, there are formulas to find the children of a node: For $j \in \{0, \ldots, B\}$, the $j$-th child of the node stored at indices $i, \ldots, i+B-1$ is stored at indices $f(i,j), \ldots, f(i,j) + B - 1$ where $f(i,j) = i(B+1) + (j+1)B$.

The code for search in a $(B+1)$-tree layout consists of a while loop that contains an inner binary search on a subarray (block) of length $B$. Very occasionally, the Btree search completes with one binary search on a block of size less than $B$. The block size, $B$, is chosen to fit neatly into one cache line. On our test machines, cache lines are 64 bytes wide, so we chose $B = 16$ for 4-byte data. Preliminary testing with other block sizes showed that this theoretically optimal choice for $B$ is also optimal in practice (see Figure 15).

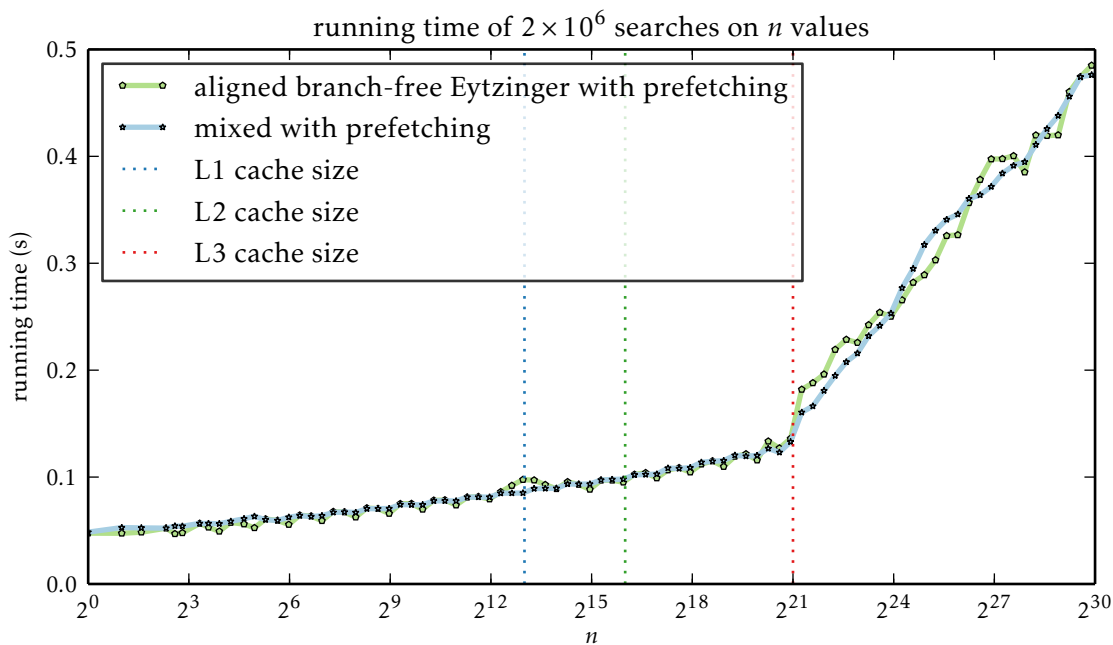

Figure 14: Performance of the branch-free mixed layout with prefetching.

### 3.3.1 Cache-Use Analysis

The height of a $(B+1)$-ary tree that stores $n$ keys is approximately $\log_{(B+1)} n$. A search in a $(B+1)$-tree visits $\log_{(B+1)} n$ nodes, each of which is contained in a single cache line, so a search requires loading only $\log_{(B+1)} n$ cache lines.

By the same reasoning as before, we expect that the top $\log_{(B+1)} C$ levels of the $(B+1)$-tree, which correspond to the first $C$ elements of the array, will be stored in the cache. Thus, the number of cache misses that occur when searching in a $(B+1)$-tree is

$$\log_{(B+1)} n - \log_{(B+1)} C = \frac{\log n - \log C}{\log(B+1)} \ .$$

Observe that this is roughly a factor of $\log(B+1)$ fewer cache misses than the $\log n - \log C$ cache misses incurred by binary search or Eytzinger search.

In our test setup, with $B = 16$, we therefore expect that the number of cache misses should be reduced by a factor of $\log 17 \approx 4.09$. When plotted, there should still be an increase in slope that occurs at $n = 2^{21}$, but it should be significantly less pronounced than in binary search.

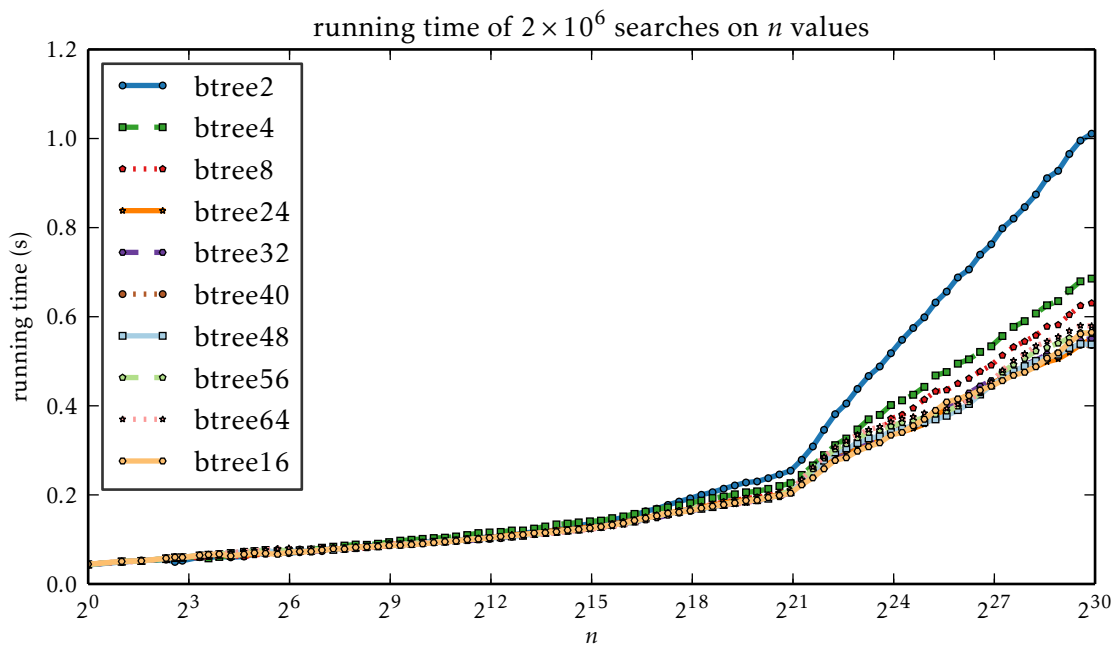

Figure 15: Performance of (branch-free) $(B+1)$-ary trees for varying values of $B$.

### 3.3.2 Naïve and Branch-Free Btree Implementations

We made and tested several implementations of the Btree search algorithm, including a completely naïve implementation, an implementation that uses C++ templates to unroll the inner search, and an implementation that uses C++ templates to generate an unrolled branch-free inner search.

Figure 16 shows the results for our three implementations of Btree search as well as branch-free binary search and our best implementation of Eytzinger search. For small values of $n$, the naïve implementation of Btree search is much slower than the unrolled implementations, and there is little difference between the two unrolled implementations. For values of $n$ larger than the L3 cache, the running time of all three Btree search implementations becomes almost indistinguishable. This is expected since, for large $n$, the running-time becomes dominated by L3 cache misses, which are the same for all three implementations.

Compared to branch-free binary search, the Btree search implementations behave as the cache-use analysis predicts: Both algorithms are fast for $n < 2^{21}$ after which both show an increase in slope. At this point, the the slope for branch-free binary search is approximately 3.8 times larger than that of the Btree search. (The analysis predicts a factor of $\log 17 \approx 4.09$.)

All three implementations of Btree search perform worse than our best Eytzinger search implementation across the entire spectrum of values of $n$. For small values of $n$, this is due to the fact that Eytzinger search is a simpler algorithm with smaller constants.

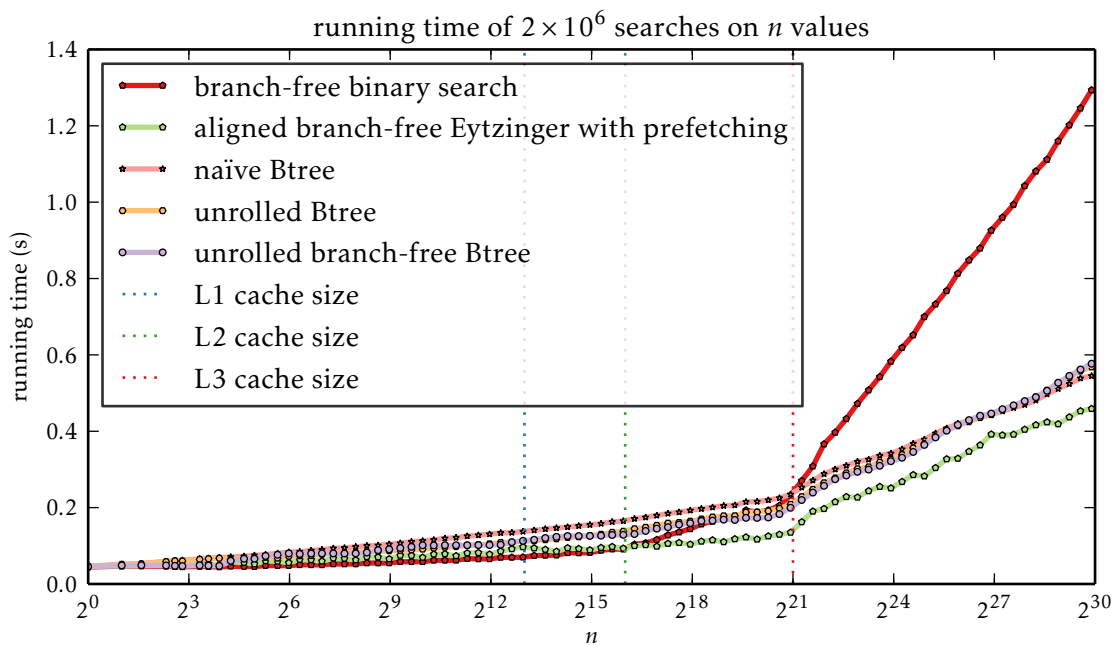

Figure 16: The performance of Btree search algorithms.

One reason Eytzinger search has smaller constants is that the inner binary search of Btrees is inherently inefficient. Our virtual Btree stores 16 keys per node, so the inner binary search has 17 possible outcomes, and therefore requires $\lceil \log_2 17 \rceil = 5$ comparisons in some cases (and in all cases for the branch-free code). This could fixed by using 16-ary trees instead, but then the number of keys stored in a block would be 15, which does not align with cache lines. We tested 16-ary trees and found them to be slower than 17-ary trees for all values of $n$.[12]

For values of $n$ larger than the L3 cache size, Eytzinger search and Btree search have the same rate of growth, but Eytzinger search starts out being faster and remains so. That the two algorithms have the same rate of growth is no coincidence: This growth is caused by RAM latency. In the case of Btrees, the time between finishing the search in one node and beginning the search in the next node is lower-bounded by the time needed to fetch the next node from RAM. In the case of Eytzinger search, the time it takes to advance from level $\ell$ in the implicit search tree to level $\ell + 4$ is lower-bounded by the time needed to fetch the node at level $\ell + 4$ from RAM. Thus, both algorithms pay one unit of latency for every 4 or 5 comparisons.

Figure 17 compares Btrees with Eytzinger search for values of $n$ smaller than the L3 cache size. One interesting aspect of this plot is that the Btree plots have a slight increase in slope at $n = 2^{16}$—at the point in which the data exceeds the L2 cache size—while the Eytzinger search algorithm shows almost no change in behaviour. This is because Btree search work synchronously with the memory subsystem; each Btree node is searched using 4–5 comparisons and then the memory subsystem begins loading the next node (from L3 cache). The Eytzinger search on the other hand begins loading the node at level $\ell + 4$ and continues to perform comparisons at levels $\ell$, $\ell + 1$, $\ell + 2$, and $\ell + 3$ while the memory subsystem asynchronously fetches the node at level $\ell + 4$. This effectively hides the L3 cache latency.

---

[12]A third alternative is to store 15 keys in blocks of size 16, but this would be outside the model we consider, in which all data resides in a single array of length $n$.

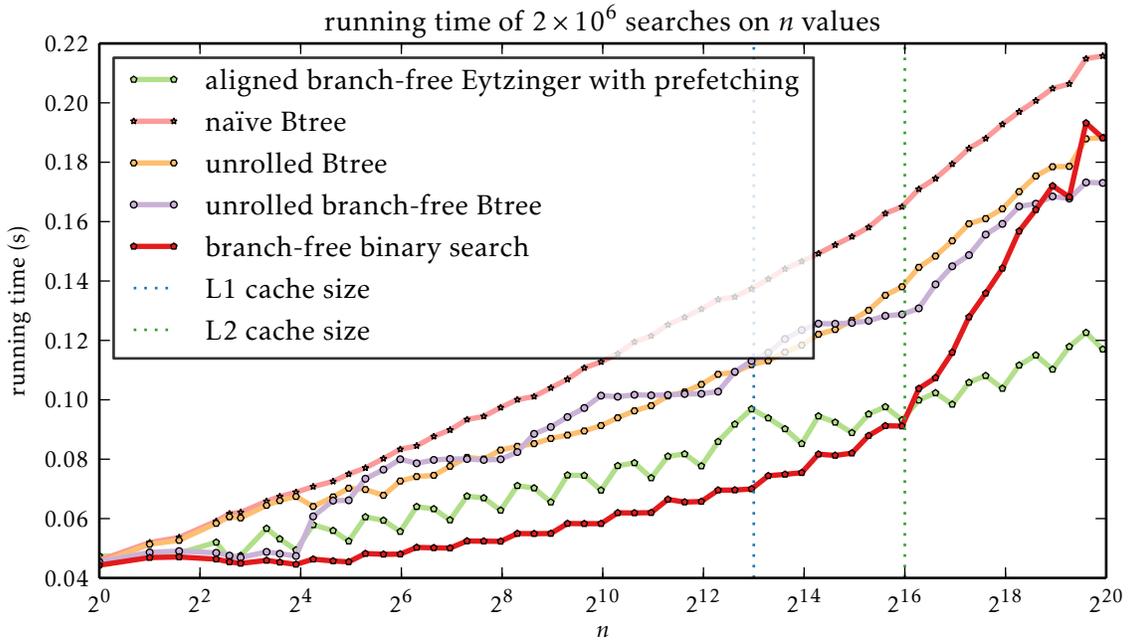

Figure 17: The performance of Btree search for small values of $n$.

**Lesson 4.** Even though it moves $B$ times more data between levels of the cache hierarchy, Eytzinger search with explicit prefetching outperforms Btrees. It does this by exploiting parallelism within the memory subsystem and between the processor and memory subsystem.

### 3.4 Van Emde Boas

Unlike the Btree layout, the Van Emde Boas layout is parameter free. It is a recursive layout that provides asympotically optimal behaviour for any cache line width, $B$. In the case where there are multiple levels of cache with different cache line widths, the van Emde Boas layout is asymptotically optimal for all levels simultaneously. Indeed, for any value of $B$, a search in a van Emde Boas layout can be viewed as searching in a sequence of binary trees whose heights are in the interval $((1/2)\log B, \log B]$, and each of these subtrees is stored in a contiguous subarray (and therefore in at most 2 cache lines). The sum of the heights of these subtrees is $\lceil \log n \rceil$.

### 3.4.1 Cache-Use Analysis

The cache-use analysis of the vEB layout approximates that of B-trees. We expect to find the most frequently accessed subtrees in the cache, which correspond, roughly, to the first $\log C$ iterations of the search algorithm. Beyond this point, the search passes through a sequence of subtrees whose height is in the interval $((1/2)\log B, \log B]$ and each of which is contained in at most two cache lines. Thus, the number of cache misses we expect to see is about $O((\log n - \log C)/\log B)$. Here, we use big-Oh notation since the exact leading constant is somewhere between 1 in the best case and 4 in the worst case.

### 3.4.2 VeB Implementations

To search in a van Emde Boas layout one needs code that can determine the indices of the left and right child of each node in the virtual binary tree. Unlike, for example, the Eytzinger layout, this is not at all trivial. We experimented with two approaches to navigating the van Emde Boas tree: machine bit manipulations and lookup tables. It very quickly became apparent that lookup tables are the faster method.

Our implementations store two or three precomputed lookup tables whose size is equal to the height of the van Emde Boas tree. In addition to these lookup tables, one must also store the path, rtl, from the root to the current node as well as an an encoding, p, of the path from the root to the current node, where each bit of p represents a left turn (0) or a right turn (1) in this path. Brodal et al. [4] reached the same conclusion when designing their implementation of the van Emde Boas layout. The most obvious (branchy) code implementing this algorithm is shown in Listing 7

In addition to the code in Listing 7, we also implemented (as much as possible) a branch-free version of the vEB code. Performance results for the vEB search implementations are shown in Figure 18, along with our best implementations of Btree search and Eytzinger arrays. As seen with other layouts, the branch-free implementation is faster until we exceed the size of the L3 cache, at which point the branchy implementation starts to catch up. However, in this case, the branchy implementation has a lot of ground to make up and only beats the branch-free implementation for $n > 2^{25}$.

The branch-free version of the vEB code is competitive with Btrees until $n$ exceeds the L3 cache size. At this point, the vEB code becomes much slower. This is because the vEB layout only minimizes the number of L3 cache misses to within a constant factor. Indeed, the subtrees of a vEB tree have sizes that are odd and will therefore never be perfectly aligned with cache boundaries. Even in the (extremely lucky) case where a vEB search goes through a sequence of subtrees of size 15, each of these subtrees is likely to be split across two cache lines, resulting in twice as many cache misses as a Btree layout.

**Lesson 5.** The vEB layout is a useful theoretical tool, and may be useful in external memory settings, but it is not useful for in-memory searching.

## 4 A Theoretical Model

There is currently no theoretical model of caches or I/O that predicts that Eytzinger search is competitive with (and certainly not faster than) Btrees. The I/O model [1] predicts that

```cpp
template<typename T, typename I>
I veb_array<T,I>::search(T x) {
    I rtl[MAX_H+1];
    I j = n;
    I i = 0;
    I p = 0;
    for (int d = 0; i < n; d++) {
        rtl[d] = i;
        if (x < a[i]) {
            p <<= 1;
            j = i;
        } else if (x > a[i]) {
            p = (p << 1) + 1;
        } else {
            return i;
        }
        i = rtl[d-s[d].h0] + s[d].m0 + (p&s[d].m0)*s[d].m1;
    }
    return j;
}
```

Listing 7: Source code for branchy vEB search.

the Eytzinger layout is a factor of $\log B$ slower than the Btree layout. However, we can augment the I/O model so that the cache line width, measured in data items, is $B$, the latency, measured in time units, for reading a cache line is $L$, and the bandwidth, measured as the number of cache lines that can be read per time unit, is $W$.[13] In this model, the time to search in a Btree is roughly

$$(L+c)\log_B n$$

where $c = O(\log B)$ is the time to search on an (in-memory) block of size $B$. This is because a B-tree search consists of $\log_B n$ rounds where each round requires loading a cache line (at a cost of $L$) and searching that cache line (at a cost of $c$).

On the other hand, if $WL > \log B$, then the memory subsystem can handle $\log B$ cache line requests in parallel without any additional overhead. In this case, the running-time of Eytzinger search with prefetching is roughly

$$\max\{L,c\}\log_B n \ .$$

(Here we assume that the local computation time for $\log B$ levels of Eytzinger search and the local computation time to do binary search on a Btree block of size $B$ is the same $c$.) This analysis shows that our experimental results do have a theoretical explanation. It also suggests two possible extensions of our experiments.

---

[13]The I/O model already has the parameter $B$, so the only real addition to the model here is the bandwidth parameter $W$, since we can take $L = 1$, without loss of generality.

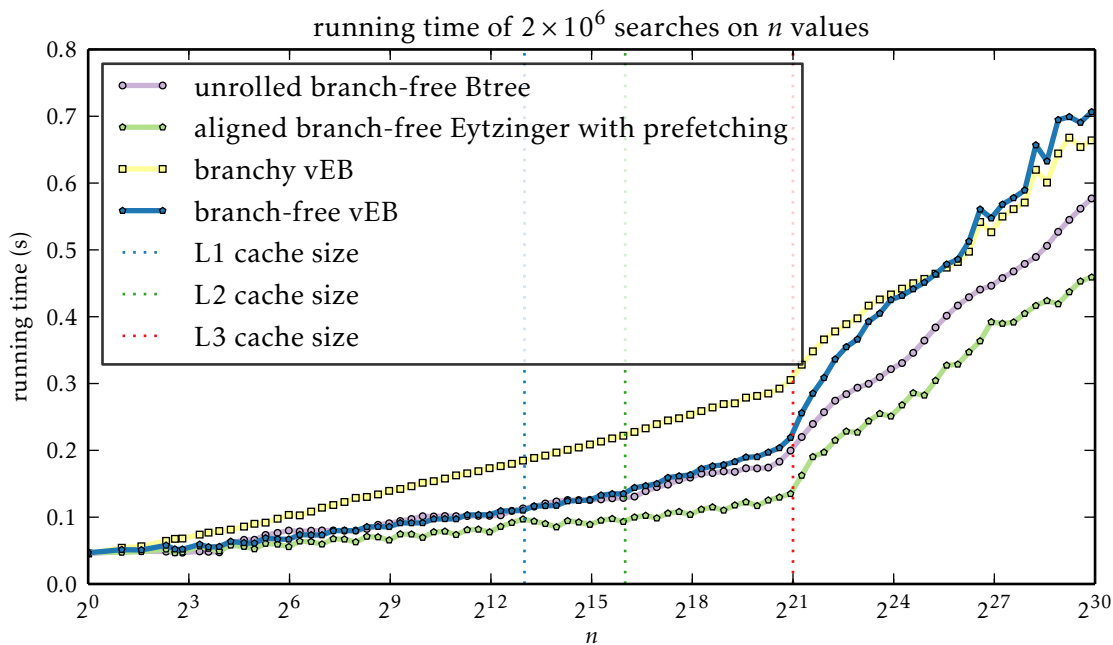

Figure 18: The performance of the vEB layout.

### 4.1 Deeper Prefetching in Eytzinger Search

If there is really an excess of bandwidth so that, for example, $WL > 2^t \log B$, then the prefetching for the Eytzinger search can be extended so, when accessing a virtual node, $v$, in the Eytzinger tree we prefetch $v$'s $2^t B$ descendants at distance $t + \log B$ from $v$ using $2^t$ prefetch instructions.

We implemented and tested this idea and found that, for the 4-byte data studied throughout this paper, it did not yield any speedup with $t = 1$ and was significantly slower with $t = 2$. This agrees with our analysis and the latency/bandwidth characteristics of our machine (Section 2.2). On our machine, $WL \approx 4.7$ and, with 4-byte data, $B = 16$, so $\log B = 4$. Thus, the Eytzinger search algorithm continuously has 4 memory requests in the memory pipeline so it is close to saturating the memory bandwidth already. Taking $t = 1$ or $t = 2$ completely saturates the memory bandwith by having 8, respectively 16, concurrent memory requests.

However, since $B$ is the number of data items that fit into a single cache line, increasing the size of data items decreases $B$. For example, with 16-byte data, $B = 4$, and $\log B = 2$. In this case, the branch-free Eytzinger search with prefetching is only prefetching two levels in advance. By choosing $t = 1$ we extend the prefetching to three levels in advance and are still only fetching 4 cache lines at any point in time.

Figure 19 shows the result of using deeper prefetching for 4-, 8-, and 16-byte data. These results agree with the predictions of the model: deeper prefetching is not beneficial for 4-byte data, but does give an improvement for 16-byte data.

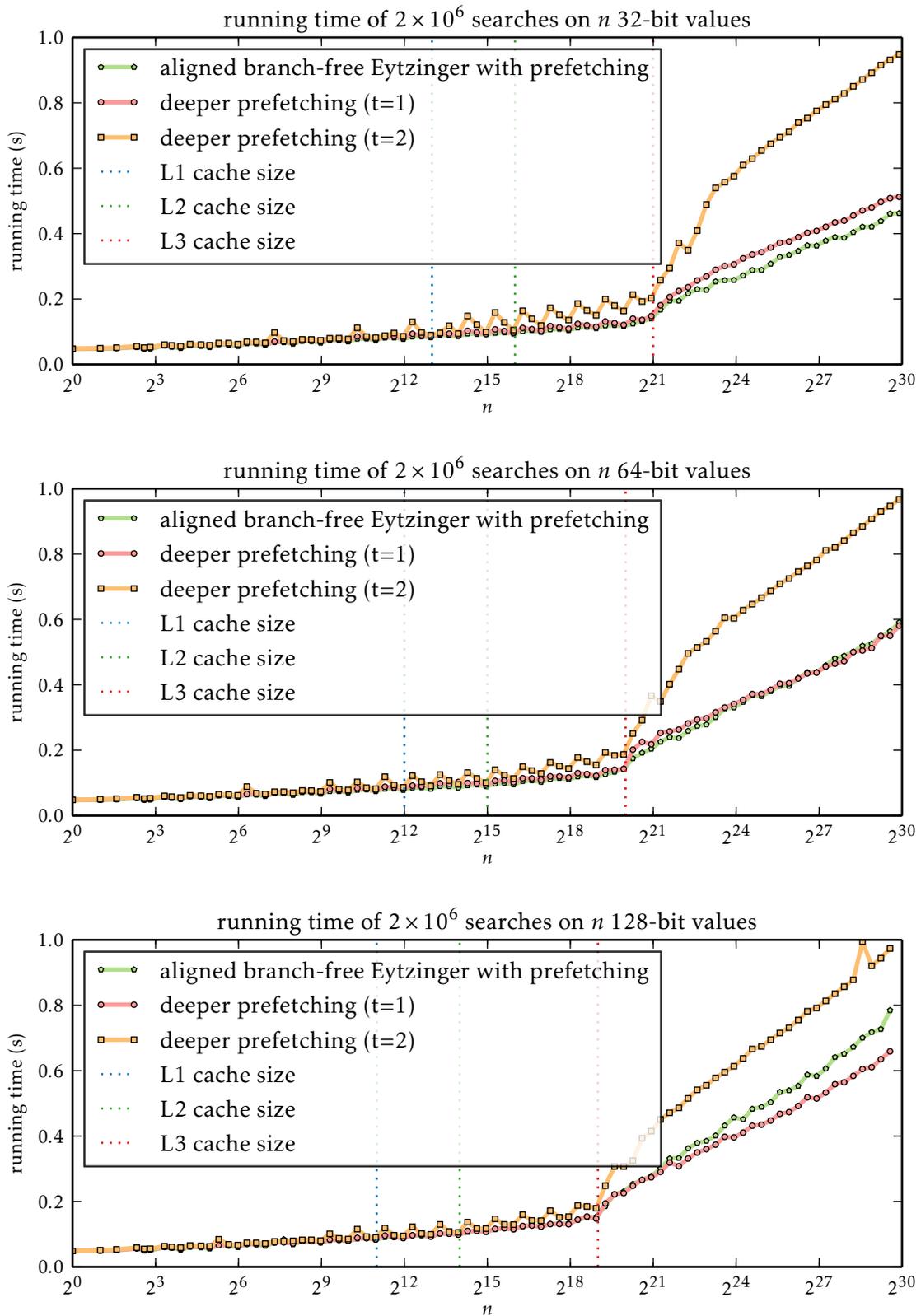

Figure 19: The effects of deeper prefetching in the Eytzinger search algorithm.

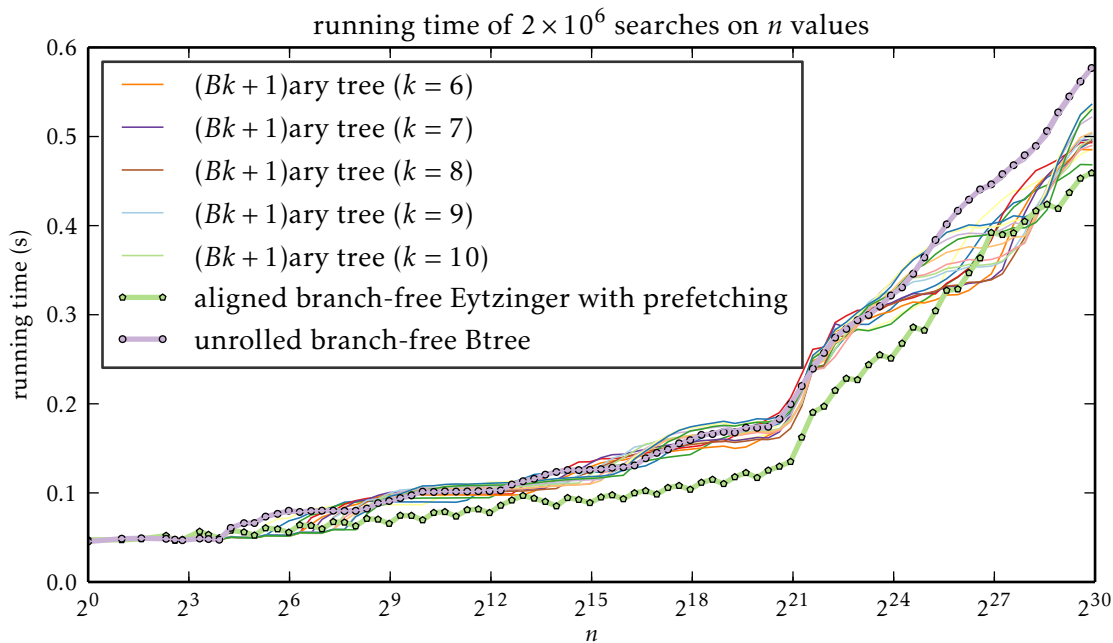

Figure 20: Testing $(Bk+1)$-ary trees.

## 4.2 The $k(B+1)$-tree

Our extended I/O model also suggests a *simulation*, in the sense that, for any $k \le W$ one can load $k$ cache lines in $L + r(k)$ time units, where $r(k)$ is the time it takes to issue $k$ prefetch requests. By doing this, one can simulate a model with cache line width $B' = kB$ and latency $L = L + r(k)$.

This immediately raises the question of whether this simulation leads to a useful data structure. To test this, we implemented a $(kB+1)$-tree where, before searching within a node, we first issue $k$ prefetch commands to load the $k$ cache lines used by that node. This idea of using prefetching to get extra-wide nodes "almost for free" is the essential idea used to speed up searches in Chen et al.'s prefetching B$^+$-tree [5]. Figure 20 shows the results of implementing this for all $k \in \{1, \ldots, 16\}$. The best choices of $k$ on this machine seem to be $k \in \{6, 7, 8, 9, 10\}$.

These results show that one can, indeed, speedup the Btree layout for some sufficiently large values of $n$ by using this technique. This offers some validation of the extension of the I/O model proposed above. For some choices of $k$ and some values of $n$, this yields a layout that is faster than the Eytzinger layout.

Unfortunately, the $(Bk+1)$-tree requires more parameter-tuning than the Eytzinger layout and only beats the Eytzinger layout for a limited range of $n$. This latter fact is because $(Bk+1)$-trees have extremely high arity; the fastest versions have internal nodes with 97 ($k = 6$) to 161 ($k = 10$) children per node. This makes them inefficient except when the bottom level is nearly full or nearly empty, i.e., when $n$ is close to a power of $(Bk+1)$. This is what causes the large oscillations that appear in Figure 20.

## 5 Further Experiments

In this section, we discuss further experiments that consider the effects of multiple threads, larger data items, data sizes that exceed the size of the translation lookaside buffer, and experiments on machines other than the Intel 4790K.

### 5.1 Multiple Threads

Since the Eytzinger search moves more data from RAM into cache, it uses more memory bandwidth. This works because, as discussed in Section 2.2, latency, not bandwidth, is the bottleneck; the bandwidth far exceeds the cache line width divided by the latency. However, when multiple threads are performing searches in parallel they could—in theory—saturate the bandwidth.

Figure 21 shows the results of tests using 2, 4, and 8 threads. In these tests, the array layout is created and $k$ threads are created, each of which performs $2 \times 10^6$ random searches (each thread seeds its own random number generator with its own unique seed). The main thread then joins each of these threads. The total time measured is the time elapsed between (before) the creation of the first thread until (after) the last thread completes.

These graphs show that, although the performance gap narrows, the Eytzinger layout continues to outperform the other layouts even with four threads simultaneously searching the array. With eight threads and very large values of $n$ the two become equally matched. We did not test more than eight threads since this seems unrealistic; the Intel 4790K has only 4 hyperthreaded cores, which appear to the operating system as 8 CPUs. It does not make sense to run more than 8 computation-intensive threads simultaneously.

### 5.2 Larger Data Items

Figure 22 shows timing results for 8-byte (64-bit) data items and 16-byte (128-bit) simulated data items. Each simulated 16-byte datum is a structure that contains an 8-byte unsigned integer and 8-bytes of padding.

In these experiments, the $(B+1)$-trees use values of $B = 8$ and $B = 4$, respectively. Similarly, the Eytzinger search algorithm prefetches the 8 virtual nodes that are 3 (for 8-byte data) and the four virtual nodes that are 2 (for 16-byte data) levels below the current node.

It is worth noting that, in both these figures, the largest value of $n$ corresponds to approximately 12GB of data. These figures show that our conclusions hold also for larger data items; the branch-free search with prefetching in the Eytzinger layout remains the method of choice. Indeed, as the size of individual data items increases, Eytzinger becomes a better and better choice.

### 5.3 Other Machines

With the exception of Figure 9, the results presented here are running times on a single machine with a single memory configuration. In order to validate these results across a wider variety of hardware, we also ran a set of experiments on a number of hardware platforms. We did this by first running experiments on machines we had access to and then by asking for help from the Internet.

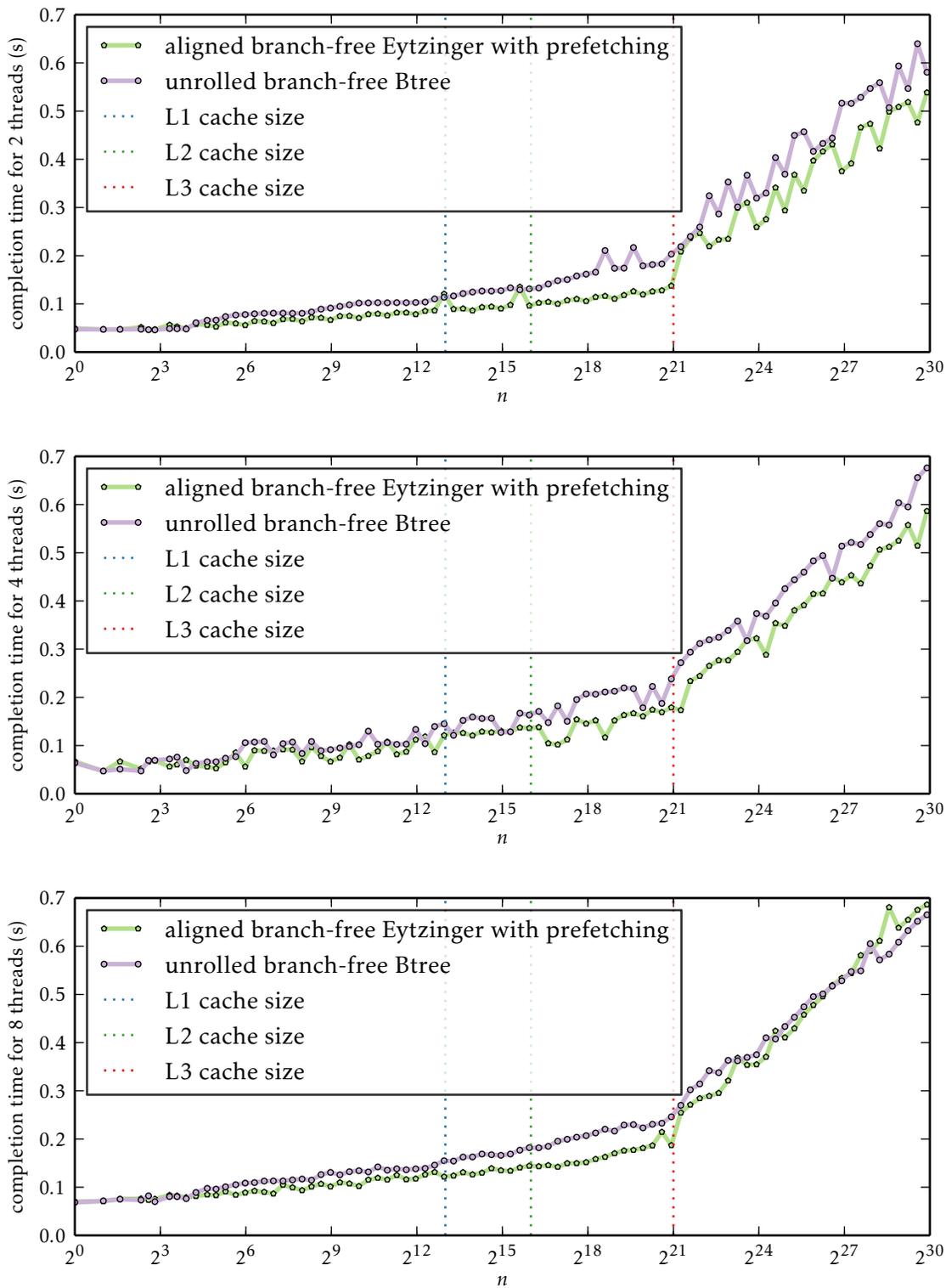

Figure 21: The performance of searches with two, four, and eight threads.

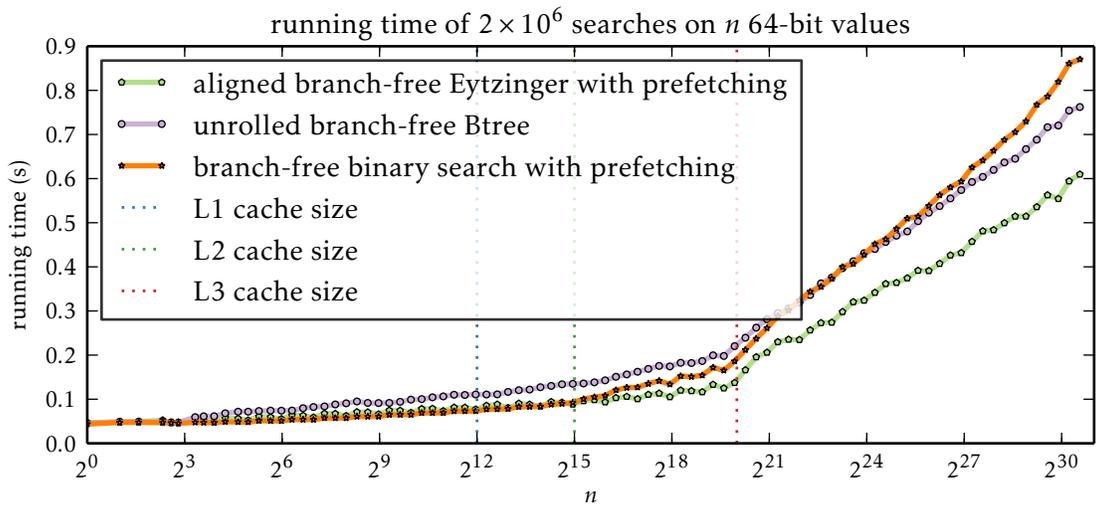

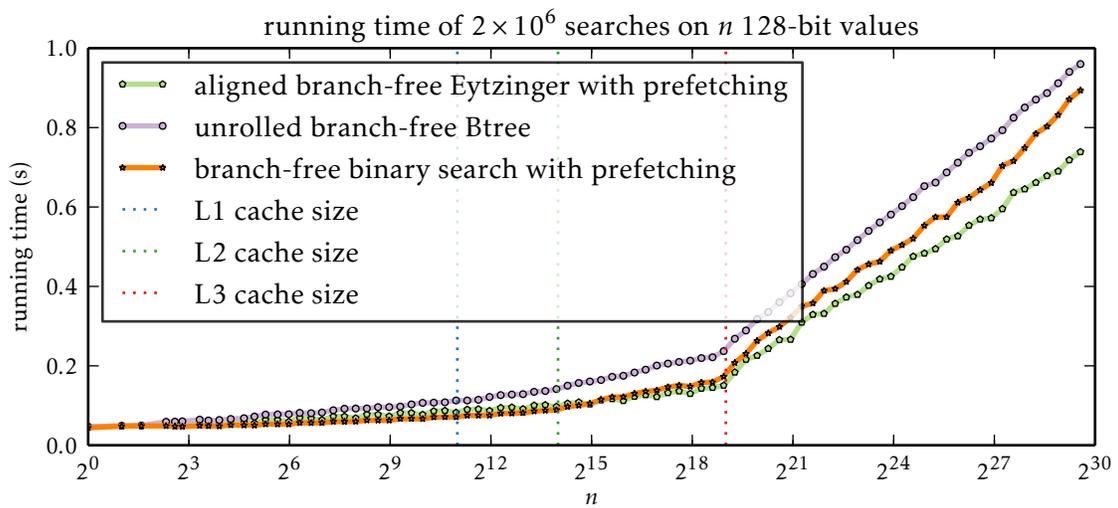

Figure 22: The performance of algorithms on 64-bit and 128-bit data.

After being surprised by the performance of the branchy Eytzinger implementation in our own initial experiments on a handful of machines, we posted to reddit asking for help testing our initial implementations on more hardware.[14] This story made the rounds of the usual hacking websites and, within two weeks we had received testing results for over 100 different CPUs. These results are available online[15] and, overwhelmingly, they showed that our branchy implementation of Eytzinger search was faster, or at least as fast as our branchy implementation of Btrees. It was then that we started micro-optimizing our code to help understand the reason for this, which led to the conclusions in the current paper.

The implementations studied in the current paper have been refined significantly since our initial round of Internet experiments, so we went back to the Internet for a second round of experiments, which we are also keeping track of online.[16] This data is still arriving but in the vast majority of cases so far, the conclusions agree with those we obtained on the Intel 4790K. There are two notable exceptions:

1. With the Atom family of processors, the branch-free code is typically faster than the branchy code over the full range of values of $n$. As discussed in Section 3.1.4, this is due to fact that the Atom processors do not perform speculative execution so the branchy code does not obtain the benefit of implicit prefetching.

2. With some processors (the AMD Phenom II, Intel Celeron J1900, and Intel E3-1230 are examples) the code that does explicit prefetching is much slower, especially for smaller values of $n$. This seems to be due to executing prefetch instructions that attempt to prefetch data outside the input array. Although the documentation for the __builtin_prefetch() instruction specifically allows prefetching invalid addresses, this seems to cause a significant performance hit on these processors. On such architectures, a work-around is to use a bit-mask to ensure that we never prefetch an address outside the input array.[17]

    Figure 23 shows the results of using this workaround on the E3-1230. Using a mask results in reliably fast performance. We also ran the same test on the Intel 4790K (see Figure 24), and the use of the mask had almost no impact on performance. Thus, for reliably fast code across the largest number of architectures using a mask to prevent out-of-bounds prefetches seems advisable.

## 6 Conclusions and Future Work

We have compared the relative performance of four memory layouts for ordered arrays and the algorithms for searching in these layouts.

---

[14]https://www.reddit.com/r/compsci/comments/35ad8d/alternatives_to_sorted_arrays_for_binary_searching/
[15]http://cglab.ca/~morin/misc/arraylayout/
[16]http://cglab.ca/~morin/misc/arraylayout-v2/
[17]Specifically, we take the bitwise AND of the prefetched index and $2^{\lfloor \log n \rfloor} - 1$.

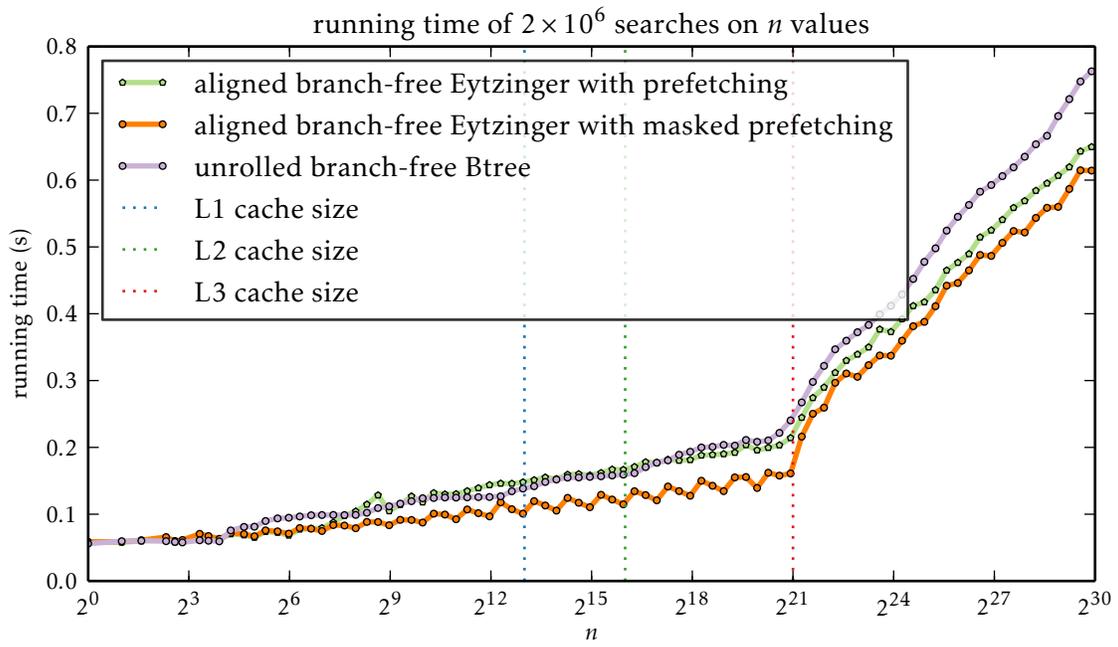

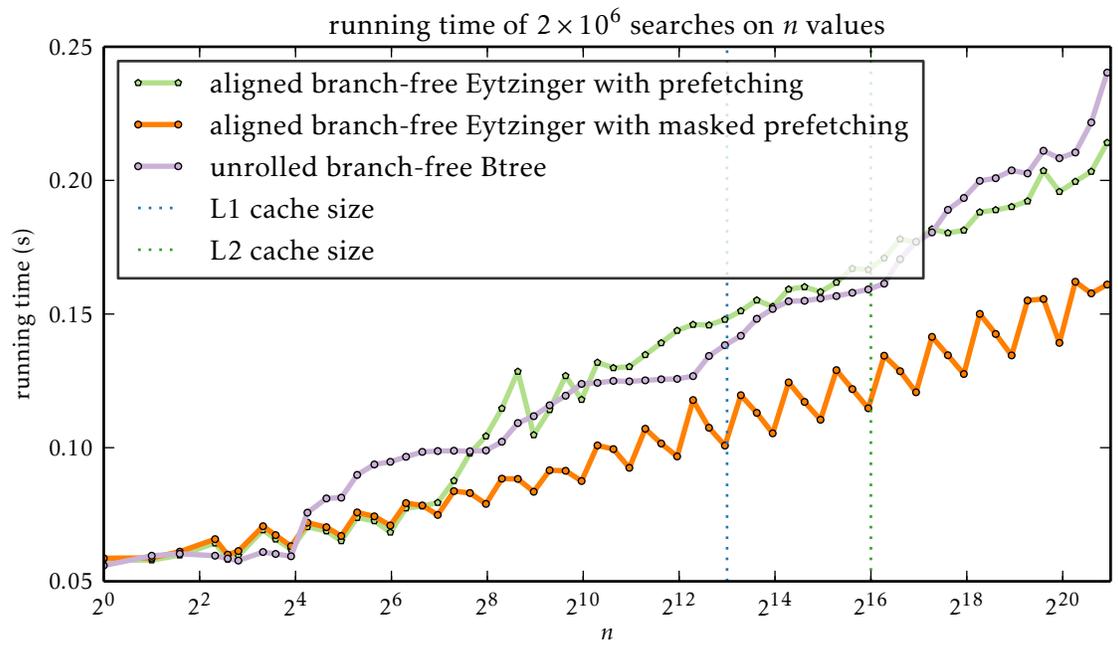

Figure 23: Using masking on the Intel E3-1230 to prevent out-of-bounds prefetches.

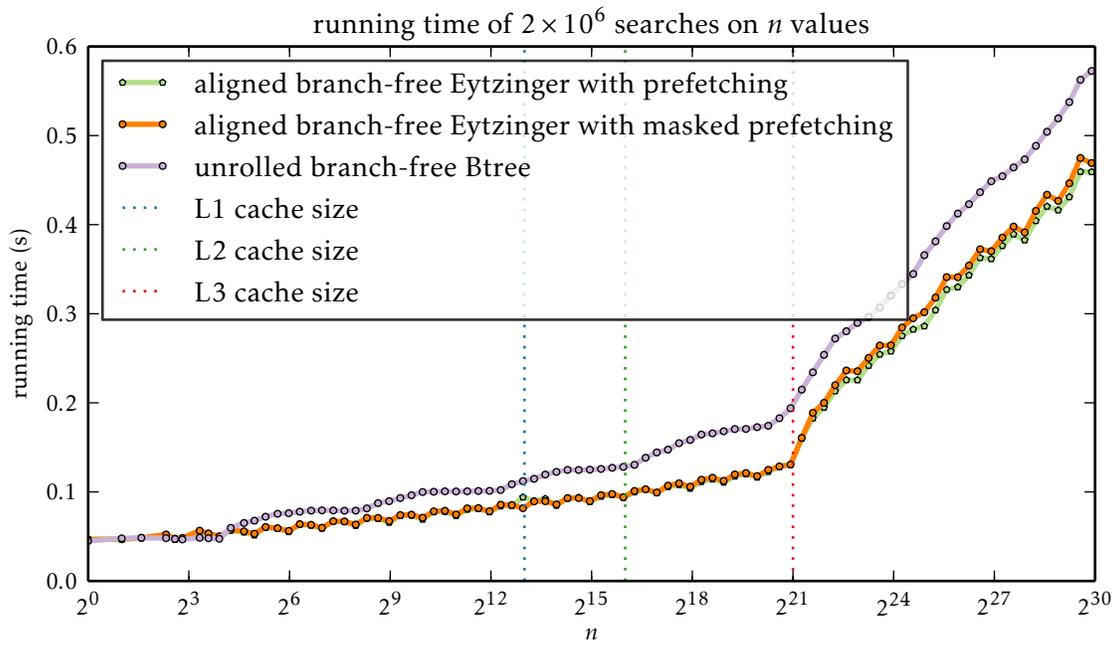

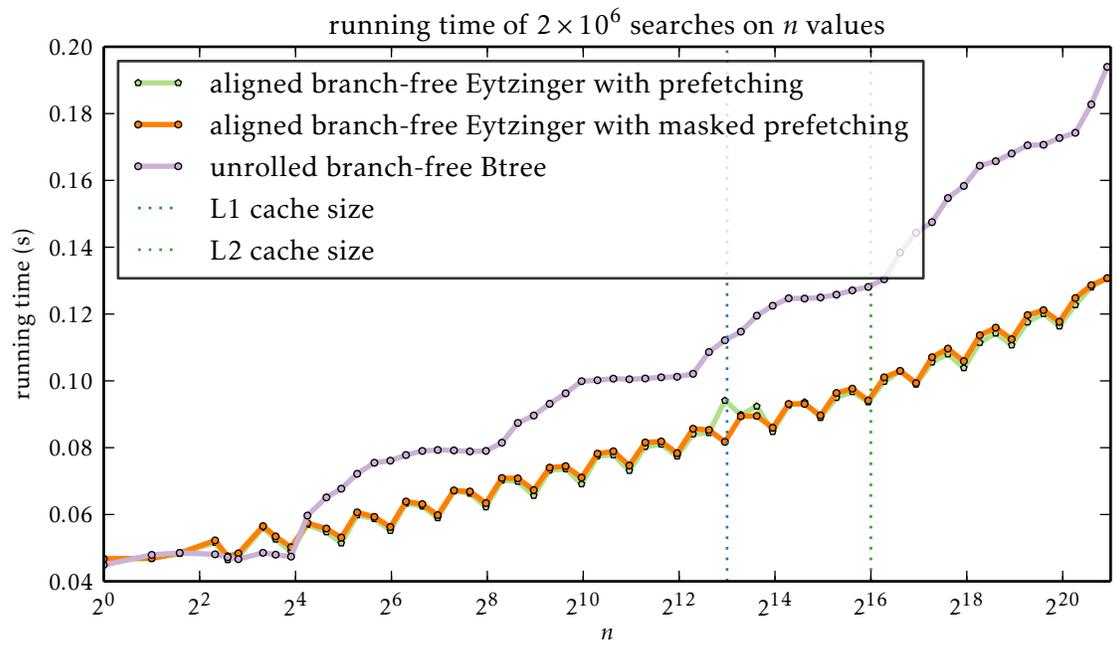

Figure 24: Masking prefetches on the Intel 4790K has negligible impact on performance.

### 6.1 Eytzinger is Best

Our conclusion is that, for the fastest speed across a wide-variety of array lengths, an Eytzinger layout and a branch-free search algorithm with explict prefetching is an excellent choice. For reliable performance across a wide variety of architectures, we recommend the variant that uses a bit mask to prevent prefetching out of bounds.

Not only is searching in an Eytzinger layout fast, it is simple and compact to implement: Counting semicolons, the Eytzinger search code consists of 5 lines of code, whereas the fastest Btree code is 19 lines plus a separate inline subroutine for the inner search. If we look at the compiled code, the Eytzinger code contains 26 assembly instructions and the Btree code contains 78.

Our conclusion is contrary to our expectations and to previous findings and goes against conventional wisdom, which states that accesses to RAM are slow, and should be minimized. A more accurate statement, that accounts for our findings, is that accesses to RAM have high latency and this latency needs to be mitigated. This was formalized in Section 4 and the resulting computational model does have some predictive power: It lead us to deeper prefetching in the Eytzinger search algorithm and the $(Bk + 1)$-tree, which behave more or less as the model predicts.

Practically speaking, this work opens up a whole class of algorithms and data structures that may be simpler to implement and faster in practice. This raises our first open question:

**Open Problem 1.** Can memory request pipelining be used to design faster algorithms for other problems?

An obvious candidate to consider in the context of Open Problem 1 is the problem of sorting where the heap-sort algorithm [7, 26], which already uses the Eytzinger layout, is an obvious starting point. Scouring the internet, we have only been able to find one implementation, due to Piotr Tarsa, of heap-sort that experiments with prefetching.[18]

### 6.2 For Big Data, Branchy Code can be a Good Prefetcher

Today, no computer science research paper is complete unless it says something about big data. For small data sets, micro-optimizations such as replacing branching with conditional instructions can produce significant improvements in running-time. For instance, our branch-free implementation of binary search is about twice as fast as the branchy implementation for $n < 2^{16}$ (refer to Figure 6).

However, as the amount of data increases beyond the size of the various processor caches, the branchy implementations become faster. Even for binary search, which represents the worst possible case for branch prediction, the branchy code eventually outperforms the branch-free code by an increasing margin (refer to Figure 7).

Our experiments show that this speedup is a result of the interaction between a long processor pipeline, speculative execution, and the memory subsystem. Indeed, spec-ulative execution acts as a form of prefetching that can speed up memory access patterns that are far too complicated for current prefetching techniques. In extreme cases (like the Eytzinger layout) the resulting prefetches are perfectly correct even four branches ahead of the current execution point.

---
[18]See the file `sortalgo/binaryheapsortaheadsimplevarianta.hpp` in https://github.com/tarsa/SortingAlgorithmsBenchmark/

In hardware design, the general idea of using speculative execution as a prefetching strategy is called *runahead execution* [17]. Our results show that, in a processor that has speculative execution, one can obtain many of the benefits of runahead even on processors that do not specifically implement it. While this may not be news to hardware experts, to the best of our knowledge, we are unaware of any algorithms (or even implementations) that deliberately make use of it.

Our results also show that one has to be very careful with micro-optimizations. Optimizations that offer great speedups on small or moderate-sized inputs can create great slowdowns on large inputs. For instance, if we compare branch-free versus branchy binary search for $n = 2^{16}$, we find that branch-free binary search is about twice as fast. However, for $n = 2^{30}$, branch-free binary search is about 45% slower.

## Acknowledgements

We are grateful to Gerth Brodal, Rolf Fagerberg, and Riko Jacob who kindly made their source code available to us. However, for a variety of reasons, most of which have to do with the 12 years that have elapsed since it was written, we found it impossible to use as a starting point for our own code.

We are also grateful to the anonymous referees for bringing references [22] and [14] to our attention.

We appreciate the efforts of Simon Gog, JEA's Reproducibility Reviewer, who worked with us through the process of obtaining a Replicated Computational Results certificate for this paper. Thanks to his efforts, the accompanying code is much more useful for reproducing our experiments.

Finally, we are also grateful to the many people who responded to our online calls for help in testing these algorithms.[19]

---

[19] http://cglab.ca/~morin/misc/arraylayout-v2/